\documentclass[10pt,preprint,showpacs,preprintnumbers,twocolumn]{revtex4}
\usepackage{amssymb,amsthm,epsfig,graphics,graphicx,subfigure}

\renewcommand{\d}{\mathrm{d}}
\renewcommand{\t}{\mathrm{Ta}}
\renewcommand{\c}{\mathrm{corr}}

\begin{document}

\title{Lifetime of Oil Drops Pressed by Buoyancy Against a Planar Interface. II. Large Drops}

\author{Clara Rojas}
\email{clararoj@gmail.com}
\author{M\'aximo Garc\'ia-Sucre}
\author{Germ\'an Urbina-Villalba}

\affiliation{Instituto Venezolano de Investigaciones Cient\'ificas (IVIC), Centro de Estudios Interdisciplinarios de la F\'isica, Apdo 20632, Caracas 1020A, Venezuela}

\date{\today}

\begin{abstract}
In a previous report [C. Rojas, G. Urbina-Villalba, M. Garc\'ia-Sucre, Phys. Rev. E {\bf 81}, 016302 (2010)] it was shown that Emulsion Stability Simulations (ESS) are able to reproduce the lifetime of micrometer-size drops of hexadecane pressed by buoyancy against a planar water/hexadecane interface. It was confirmed that small drops ($r_i < 10\,\mu$m) stabilized with $\beta\mathrm{-casein}$ behave as non-deformable particles, moving with a combination of Stokes and Taylor tensors as they approach the interface. Here, a similar methodology is used to parametrize the potential of interaction of drops of soybean oil stabilized with Bovine Serum Albumin (BSA). The potential obtained is then employed to study the lifetime of deformable drops in the range $10\,\mu\mathrm{m} \leq r_i \leq 1000\,\mu\mathrm{m}$. It is established that the average lifetime of these drops can be adequately replicated using the model of truncated spheres. However, the results depend sensibly on the expressions of the initial distance of deformation and the maximum film radius used in the calculations. The set of equations adequate for large drops is not satisfactory for medium-size drops ($10\,\mu$m $\leq r_i \leq 100\,\mu$m.) and vice versa. In the case of large particles, the increase of the interfacial area as a consequence of the deformation of the drops generates a very large repulsive barrier which opposes coalescence. Nevertheless, the buoyancy force prevails. As a consequence, it is the hydrodynamic tensor of the drops which determine the characteristic behavior of the lifetime as a function of the particle size.  While the average values of the coalescence time of the drops can be justified by the mechanism of film thinning, the scattering of the experimental data of large drops cannot be rationalized using the methodology previously described. A possible explanation of this phenomenon required elaborate simulations which combine deformable drops, capillary waves,  repulsive interaction forces, and a time-dependent surfactant adsorption. 
 
\end{abstract}


\maketitle

\section{Introduction}
Proteins are polyelectrolytes with a complex structure, whose spatial conformations strongly depend on the pH of the dispersing medium. They are commonly used as stabilizers in food emulsions, although it is often difficult to justify the stability of these emulsions in terms of the molecular properties of the proteins. The adsorption of these molecules to the oil/water interface is usually slow, markedly affecting the viscoelastic properties of the interface and the coalescence time between the drops. 

It is not yet possible to measure the coalescence time between two oil drops of micrometer size suspended in quiescent media in the absence of external forces. However, the diffusion tensor of the drops depends on the applied force. In one approximation to the problem, experimental measurements of the coalescence time between a drop and a planar interface are used. Proteins are expected to favor the occurrence of tangentially immobile interfaces, due to their high molecular weight. Dickinson {\it et al.} \cite{dickinson:1988} showed that the lifetime of the drops of hexadecane stabilized with $\beta$-casein, $\kappa$-casein and lysozyme decreases with the radius of the drops ($1\mu$m $\leq r_i \leq 5\,\mu$m) when they are pressed by buoyancy against the interface. These findings were successfully explained by Basheva {\it et al.} \cite{basheva:1999} arguing that when a drop is released in the bulk of a liquid, it moves according to Stokes law until it approaches the interface to a sufficiently small distance. Then it markedly decelerates due to the increase of the viscous friction in the remaining gap. If the drop retains its spherical shape while approaching the interface, its velocity can be expressed as a combination of the diffusion tensors of Stokes and Taylor, respectively:

\begin{equation}
\label{st+ta}
 \frac{1}{v}=\frac{1}{v_\mathrm{St}}+\frac{1}{v_\t},
\end{equation}
where $v_\mathrm{St}=F/6\pi\eta r_i$ refers to the Stokes law for motion of a sphere in an unbounded liquid, and $v_\t$ is given by: 

\begin{equation}
\label{sphere-flat}
 v_{\t} = \frac{hF}{6\pi\eta r_i^2}.
\end{equation}

In Eq. (\ref{sphere-flat}) $h$ is the closest distance of approach between two spheres, $\eta$ is the dynamic viscosity of the external liquid,  $r_i$ is the radius of the small droplet, and $F$ is the external driving force:

\begin{equation}
\label{bouyancy}
F =\frac{4}{3}\pi r_i^3 \Delta\rho g,
\end{equation}
where $\Delta\rho$ is the density difference, and $g$ is the gravity.

The time elapsed from the moment in which the drop starts to move slowly, until it coalesces with the large homophase, is equal to: 

\begin{equation}
\label{lifetime}
 \tau=\int^{h_{\mathrm{ini}}}_{h_\mathrm{crit}} \frac{\d h}{v(h)},
\end{equation}
where $v(h)$ is the velocity of thinning of the liquid film between the drop and the interface, $h$ is the closest distance of approach between their surfaces, $h_\mathrm{ini}$ is the gap width at which the thinning begins, and $h_\mathrm{crit}$, is the minimum distance that can be attained (critical thickness of rupture) before the film breaks and coalescence occurs.  

Substituting Eq. (\ref{st+ta}) into Eq. (\ref{lifetime}), an inverse dependence between the lifetime of the drop at the planar interface, and its radius is obtained:

\begin{equation}
\label{Stokes-Taylor}
 \tau=\frac{9\eta}{2\Delta\rho g}\frac{1}{r_i}\left[\log\left(\frac{h_\mathrm{ini}}{h_\mathrm{crit}}\right)+\frac{h_\mathrm{ini}-h_\mathrm{crit}}{r_i}\right].
\end{equation}

Recently, our group designed a methodology that proved successful in reproducing the experimental data of Dickinson {\it et al.}. Among other results, it was found that: (a) micrometer-size drops behave as non-deformable particles which move with a combination of Stokes and Taylor tensors as they approach the interface; (b) The coalescence time of the simulations can be fitted with high accuracy to the equation: $\tau=1/(A+B\,r_i)$, where $A=-0.054103$ s$^{-1}$, $B=0.075662$ ($\mu$m$\,$s)$^{-1}$ ($r^2=0.9997$). An approximate power-law dependence indicates that: $\tau \propto  r_i^{-1.39}$ ($r^2=0.9949$) for this size range; (c) the potential of interaction has a significant influence on shape of $\tau$ vs. $r_i$. The experimental data can only be accurately reproduced assuming negligible repulsive barriers; and (d) for small drops the uncertainty in the exact form of the total potential of interaction between a drop and the interface in the presence of proteins mostly resides in the steric contribution. This part of the potential depends on the Flory-Huggins parameter, the interfacial area of the protein, and the thickness of the protein layer which lies above the interfacial boundary.    

In order to replicate the lifetime of larger drops ($10\,\mu\mathrm{m}-1000\,\mu\mathrm{m}$) published by Basheva {\it et al.} \cite{basheva:1999}, we had taken advantage of the dependence of the coalescence time on the steric potential. First, we parametrized the steric potential of micrometer-size drops by reproducing the their experimental lifetime using the tensors of Stokes and Taylor. Then, we investigated the dependence of $\tau$ vs. $r_i$ for bigger drops employing the parameters of the potentials previously selected. 

Unlike small drops, large drops are expected to deform as they approach the interface. According to Reynolds \cite{landau:1984} the velocity of drainage of a thin liquid film between two circular sections can be approximated by:

\begin{equation}
\label{reynolds}
V_\mathrm{Re}=\frac{2Fh^3}{3\pi\eta r_\mathrm{f}^4},
\end{equation}
where $r_\mathrm{f}$ is the radius of the film. This radius can be found from the stress balance of the film at the planar surface \cite{basheva:1999}. It increases with the radius of the drop and inversely proportional to the interfacial tension ($\gamma$): 

\begin{equation}
\label{rf}
r_\mathrm{f}^2=\frac{F r_i}{\pi\gamma}.
\end{equation}
Substituting Eqs. (\ref{reynolds}) and  (\ref{rf}) into Eq. (\ref{lifetime}) an expression for the lifetime of deformable droplets is obtained \cite{basheva:1999}:

\begin{equation}
\label{tau}
\tau=\frac{3\eta F r_i^2}{4\pi\gamma^2}\left(\frac{1}{h_\mathrm{c}^2}-\frac{1}{h_\mathrm{in}^2}\right) .
\end{equation}

It is clear from Eq. (\ref{tau}) that in the case of large drops, the lifetime increases as a function of a particle radius. However, the force usually depends on the particle radius, and the critical distance of rupture depends on the radius, the interfacial tension, and the force. An estimation of the coalescence time can be obtained discarding the second term of Eq. (\ref{tau}), and using the equation proposed by Vrij \cite{vrij:1968,vrij:1966} for the critical thickness of rupture. In this case, $\tau \propto  r_i^{(25/7)}$.

This article is structured as follows: In Sec. \ref{emulsion} an overview of Emulsion Stability Simulations is presented. In Sec. \ref{computational} the computational details of the simulations are given and the parametrization procedure explained. In Sec. \ref{results} the results are shown and in the final section  \ref{conclusions} the conclusion is given.

\section{Emulsion Stability Simulations}
\label{emulsion}

A detailed description of the algorithm of Emulsion Stability Simulations can be found in \cite{urbina:2000,urbina:2004,urbina:2009a,toro:2010,rojas:2010a} and references therein. In ESS the particles move with an equation of motion similar to the one of Brownian Dynamic Simulations \cite{ermak:1978}:

\begin{equation}
\label{brownian}
\mathbf{r}_{\mathrm{p},i}(t+\Delta t)=\mathbf{r}_{\mathrm{p},i}(t)+\frac{D_{i}\mathbf{F}_i}{k_B T}\Delta t+\mathbf{R},
\end{equation}
where $\mathbf{r}_{\mathrm{p},i}$ is the position of particle $i$, $D_i$ is the diffusion constant, $\mathbf{F}_i$ is the total force acting on $i$, $k_B$ is the Boltzmann constant, $T$ the temperature, $\Delta t$ the time step, and $\mathbf{R}$ a random term which represents the Brownian motion of the particle.  The diffusion constant is equal to $D_i=D_0 f_\c^{(1)}f_\c^{(2)}$ where $D_0=k_BT/6\pi\eta r_i$. The first correction term, $f_\c^{(1)}$, takes into account those factors that change the diffusion constant of a particle at infinity dilution. The second correction term $f_\c^{(2)}$  takes into account the hydrodynamic interactions between the particles, caused by the movement of the surrounding liquid as the particles diffuse \cite{urbina:2009a}.

A calculation begins distributing a set of oil drops in a cubic box of side length $L$. It is assumed that the molecules of oil mainly determine the van der Waals interaction between the particles. Instead, the repulsive interactions depend on the amount and chemical nature of the surfactant molecules adsorbed to the interface of the drops. The program has several routines for apportioning surfactant molecules amongst the drops. In the most simple case, the surfactant is distributed evenly and instantaneously between the available interfaces. Once the surfactant has been allocated, the surface properties of the drops (like charge, interfacial tension, etc.) can be computed. Following, the diffusion constant and interaction forces can be calculated and the drops moved according to Eq. (\ref{brownian}). At every time step, the program checks for the coalescence of drops. In the case of non-deformable drops, coalescence occurs whenever the distance of separation between the centers of mass of the drops, $r_{ij}$, is smaller than the sum of their radii. When this happens, the former drops disappear and a new bigger drop is created at the center of mass of the colliding particles. The radius of the new drop results from the conservation of volume.

The present version of the code can simulate the behavior of systems of non-deformable or deformable drops. In both cases the particles follow the same equation of motion (Eq. (\ref{brownian})) but the analytical form of the diffusion tensors and the interaction forces change. 

If the mode of deformable droplets is selected, it is assumed that the deformation occurs independently of the energy required for this process. The model of truncated spheres is used to simulate the change of shape of the drops \cite{ivanov:1999,danov:1993b,danov:1993a}. According to this model, three regions of approach can be defined:

\begin{description}
\item{Region I:} The distance of separation between the centers of mass of the drops, $r_{ij}$, is larger than $r_i+r_j+h_{\mathrm{ini}}$, where $h_\mathrm{ini}$ stands for the initial distance of deformation of the drops. Hence, the drops behave as spherical particles.  

\item{Region II:} This region covers the range of distances between the beginning of the deformation $r_\mathrm{f}\neq 0$, and the attainment of the maximum film radius: $r_\mathrm{f}=r_{\mathrm{fmax}}$. As soon as the drops enter Region II, they change their shape from spheres to truncated spheroids. Following the model proposed by Danov {\it et al.} \cite{danov:1993b,danov:1993a} the closest distance of separation between the surfaces of the drops is assumed to be constant $h=h_\mathrm{ini}$ \cite{ivanov:1999} while the radius of the film increases:

\begin{eqnarray}
\label{def1}
\nonumber
 h_\mathrm{ini}&+&\left(\sqrt{r_i^2-r_\mathrm{fmax}^2}+\sqrt{r_j^2-r_\mathrm{fmax}^2}\right)\\
&<& r_{ij}< r_i+r_j+h_{\mathrm{ini}},\\
\label{def2}
 r_\mathrm{f}&=&\sqrt{r_i^2-\left[\frac{r_i(r_{ij}-h_\mathrm{ini})}{r_i+r_j}\right]^2}.
\end{eqnarray}

\item{Region III:} The maximum film radius has been attained, $r_\mathrm{f}=r_{\mathrm{fmax}}$, and the intervening liquid between the drops drains until it reaches a critical distance of rupture:

\begin{eqnarray}
\label{def3}
h_\mathrm{crit}&+&\left(\sqrt{r_i^2-r_\mathrm{fmax}^2}+\sqrt{r_j^2-r_\mathrm{fmax}^2}\right)\\
\nonumber
&<& r_{ij}\\
\nonumber
&<& h_\mathrm{ini}+\left(\sqrt{r_i^2-r_\mathrm{fmax}^2}+\sqrt{r_j^2-r_\mathrm{fmax}^2}\right),\\
\label{def4}
 h&=&r_{ij}-\left(\sqrt{r_i^2-r_\mathrm{fmax}^2}+\sqrt{r_j^2-r_\mathrm{fmax}^2}\right).
\end{eqnarray}
\end{description}

Accurate estimation of the initial distance of deformation $h_\mathrm{ini}$, is very difficult since it results from a balance between hydrodynamic and interaction forces. In the present calculations the soundness of two different expressions for $h_\mathrm{ini}$ was compared. The first equation was obtained fitting the curves of $h_\mathrm{ini}(r_{i},\gamma)$ -given in Ref. \cite{danov:1993a}- with a polynomial expression:

\begin{eqnarray}
\label{hini_program}
\nonumber
 h_\mathrm{ini}&=&\left[1.2932\times 10^8-8.6475\times 10^{-9}\right.\\
\nonumber
 &\times&\left.\exp(-r_i/1.8222\times 10^{-6}\right]\\
 &\times&\frac{3.3253+5.9804\exp(-\gamma/0.00402)}{3.3253+5.9804\exp(-10^{-3}/0.00402)}.
\end{eqnarray}

When the above expression is used to estimate $h_\mathrm{ini}$, the maximum film radius is approximated by:

\begin{equation}
 \label{rmax}
r_\mathrm{fmax}=\sqrt{r_ih_\mathrm{ini}}.
\end{equation}

As an alternative formula for $h_\mathrm{ini}$, we considered a fraction ($f_h$) of the expression obtained using the lubrication approximation in the presence of the buoyancy force \cite{ivanov:1985}:

\begin{equation}
\label{hini_bouyancy}
 h_\mathrm{ini}=f_h \frac{2\, r_i^3\Delta\rho g}{3\gamma},
\end{equation}
where $f_h$ is a real number between zero and one. 
Whenever Eq. (\ref{hini_bouyancy}) is used, the maximum film radius is estimated as a fraction ($f_r$) of the value predicted by Eq. (\ref{rf}):

\begin{equation}
 \label{fr}
r_\mathrm{fmax}=f_r r_i^2\sqrt{\frac{2 g\Delta\rho}{3\gamma}}.
\end{equation}

When either an arbitrary estimation of $h_\mathrm{ini}$ is used as an input of the simulation, or an alternative equation like Eq. (\ref{hini_bouyancy}) is used for its evaluation, Eq. (\ref{def2}) cannot be employed to estimate  $r_\mathrm{f}$. For this purpose an alternative expression is implemented:
 
\begin{equation}
 \label{rfb}
r_\mathrm{f}^2=r_i^2-\left[\frac{\left(r_{ij}-h_\mathrm{ini}\right)^2-\left(r_j^2-r_i^2\right)}{2\left(r_{ij}-h_\mathrm{ini}\right)}\right]^2.
\end{equation}

Eq. (\ref{rfb}) can be deduced assuming that the film formed between two drops has a uniform thickness. As a consequence, the radius of the deformation in each drop is the same, and:

\begin{equation}
\label{hrdd}
 h = r_{ij} - d_i - d_j,
\end{equation}
where $d_k$ is the distance between the center of the truncated sphere $k$ and the center of its planar interface. Notice that when Eqs. (\ref{rfb}) and (\ref{hrdd}) are used, the surface-to-surface separation between the drops ($h$) changes during the growth of the film radius (Region II).

In regard to $h_\mathrm{crit}$, the expression published by Scheludko and others was used \cite{scheludko:1967,ivanov:1970,manev:2005a,manev:2005b}:

\begin{equation}
h_\mathrm{crit}=\left(\frac{A_H A_\mathrm{crit}}{128\gamma}\right)^{1/4}, 
\end{equation}
where $A_\mathrm{crit}= r_\mathrm{f}/10$ and $A_H$ is the Hamaker constant.

Even in the mode of deformable drops, the particles behave as spheres if $r_{ij}>r_i+r_j+h_\mathrm{ini}$. This means that the potential of interaction and the diffusion constant correspond to the ones of spherical particles within Region I. At $h=h_\mathrm{ini}$, the code calculates the dimensions of truncated spheres which are compatible with the actual distance of separation between the centers of mass of the spherical drops ($r_{ij}<r_i+r_j+h_\mathrm{ini}$). In this case, the expressions of the potentials corresponding to two truncated spheres are employed to move the drops (see Table \ref{potentials} and Ref. \cite{danov:1993a}). Those potentials are expressed in terms of the thickness of the film and its radius. The use of three regions of approach allows to develop equations which relate $r_{ij}$ to those variables. As a result, the potentials can be differentiated algebraically, and the force between the particles can be calculated (see Refs. \cite{rojas:2010a,toro:2010} for details). 

\begin{table*}[htbp]
\caption{Potentials used in the simulations. In these equations, $r_i$ is the radius of the small droplet, $r_j$ is the radius of the large drop and $h$ is the minimum distance between their surfaces.  For the van der Waals (vdW) potential: $A_H$ is the Hamaker constant, $x=\frac{h}{2r_i}$, $y=\frac{r_i}{r_j}$, $l=h+r_i+\sqrt{r_i^2-r_\mathrm{f}^2}$, $L=r_j+\sqrt{r_j^2-r_\mathrm{f}^2}$, $d=\sqrt{h^2+4 r_\mathrm{f}^2}$, $h$ and $r_\mathrm{f}$ are the thickness and radius of the film, respectively. For the electrostatic potential (elect): $\kappa^2=\frac{8\pi e^2 z^2}{\epsilon k_BT}C_\mathrm{el}$, $z$ is the charge number, $\epsilon$ is the dielectric permittivity of the medium, $C_\mathrm{el}$ is the electrolyte concentration, $e$ is the electron charge, $k_BT$ is the thermal energy, $\Psi_\mathrm{si}$ and $\Psi_\mathrm{sj}$ are the surface potentials for the small and large drops, respectively. For the steric potential (st): $V_\mathrm{w}$ is the molar volume of the solvent, $\chi$ is the Flory-Huggins solvency parameter of the protein, $\bar{\phi}_j$ and $\bar{\phi}_i$ are the average volume fraction of the protein around each sphere, $\bar{\phi}_i=\frac{3r_i^2\Gamma M_\mathrm{p}}{\rho_\mathrm{p} N_A \left[(r_i+\delta)^3-r_i^3\right]}$, with $\Gamma$ the number of molecules per unit area, $\rho_\mathrm{p}$ the density of the protein, $M_\mathrm{p}$ the molecular weight of the protein and $\delta$ the width of the protein layer. Volumes $v_a$, $v_b$, and $v_c$ depend on $h$. Their explicit geometrical expressions can be seen in Ref. \cite{lozsan:2006}. For the dilatational (extensional) potential (dil):  $\gamma_0$ is the interfacial tension  and $r_a=\frac{2r_i r_j}{r_i+r_j}$. For the bending potential (bend): $|B_0|=1.6\times 10^{-12}\,$N \cite{kralchevsky:1991a, kralchevsky:1991b} is the interfacial bending moment.}
\bigskip
\begin{tabular}{c|l}
\hline\hline
\raisebox{1.0ex}{}\\
&\raisebox{-1.0ex}{$V_\mathrm{vdW}=-\frac{A_H}{12}\left[\frac{y}{x^2+xy+x}+\frac{y}{x^2+xy+x+y}+2 \ln\left(\frac{x^2+xy+x}{x^2+xy+x+y}\right)\right].$ \hspace{3.4cm} \cite{hamaker:1937}}\\
&\raisebox{-3.0ex}{$V_\mathrm{elect}=\frac{64\pi}{\kappa}C_\mathrm{el}k_BT\tanh\left(\frac{e\Psi_\mathrm{si}}{4k_BT}\right)\tanh\left(\frac{e\Psi_\mathrm{sj}}{4k_BT}\right)\times e^{-kh}\left[\frac{2r_ir_j}{\kappa(r_i+r_j)}\right].$ \hspace{2.75cm}\cite{danov:1993a}}\\
\raisebox{-0.0ex}{Spherical}
&\raisebox{-3.0ex}{$V_\mathrm{st}=\frac{4k_BT}{3V_1}\bar{\phi}_i\bar{\phi}_j\left(\frac{1}{2}-\chi\right)\left(\delta-\frac{h}{2}\right)^2\left[\frac{3(r_i+r_j)}{2}+2\delta+\frac{h}{2}-\frac{3(r_j-r_i)^2}{2(h+r_i+r_j)}\right],$ \hspace{2.5cm}\cite{lozsan:2005,lozsan:2006}}\\
&\raisebox{-1.0ex}{\hspace{9cm} $\delta < h < 2\delta$.}\\
&\raisebox{-3.0ex}{$V_\mathrm{st}=\frac{k_BT}{V_1}\left(\frac{1}{2}-\chi\right)\left[\left(\bar{\phi}_j\right)^2\left(\frac{v_a^2}{v_c}-v_a\right)+\left(\bar{\phi}_i\right)^2\left(\frac{v_b^2}{v_c}-v_b\right)+\,2\bar{\phi}_i\bar{\phi}_j\left(\frac{v_av_b}{v_c}\right)\right],$ \hspace{1.65cm}\cite{lozsan:2006}}\\
&\raisebox{-1.0ex}{\hspace{9cm} $0 < h < \delta$.}\\
\raisebox{1.0ex}{}\\
\hline\hline 
&\raisebox{-4.0ex}{$V_\mathrm{vdW}=-\frac{A_H}{12}\left\{\frac{2r_j(l-h)}{l(L+h)}+\frac{2r_j(l-h)}{h(l+L)}\right.+2\ln\left[\frac{h(l+L)}{l(h+L)}\right]+\frac{r_\mathrm{f}^2}{h^2}$} \\
&$\hspace{1.1cm} -\frac{l-h}{L}\frac{2r_\mathrm{f}^2}{hl}-\frac{l-r_i-(L-r_j)}{2l-2r_i-h}\frac{2r_\mathrm{f}^2}{hl}-\frac{2(L-r_j)-h}{2l-2r_i}\frac{d-h}{2h}+\frac{2r_jL^2(l-h)}{hl(l+L)(L+h)}$\\
&$ \hspace{1.1cm} -\frac{2r_j^2}{h(2l-2r_i-h)}\frac{l^2+r_\mathrm{f}^2}{(l+L)(l+L-2r_j)}+\frac{2r_j^2d}{(2l-2r_i-h)\left[(h+L)(h+L-2r_j)-(l-h)(l-2r_i-h)\right].}$\\
&$\hspace{1.1cm}\left.-\frac{4r_j^3(l-h)}{(l+L)(l+L-2r_j)\left[(h+L)(h+L-2r_j)-(l-h)(l-2r_i-h)\right]}\right\}.$ \hspace{3.9cm} \cite{danov:1993a}\\
\\
&\raisebox{-1.0ex}{$V_\mathrm{elect}=\frac{64\pi}{\kappa}C_\mathrm{el}k_BT\tanh\left(\frac{e\Psi_\mathrm{si}}{4k_BT}\right)\tanh\left(\frac{e\Psi_\mathrm{sj}}{4k_BT}\right)\times e^{-kh}\left[r_\mathrm{f}^2+\frac{2r_ir_j}{\kappa(r_i+r_j)}\right].$ \hspace{2.16cm}\cite{danov:1993a}}\\
\raisebox{-3.0ex}{Deformable}
&\raisebox{-3.0ex}{$V_\mathrm{st}=\frac{4k_BT}{3V_1}\bar{\phi}_i\bar{\phi}_j\left(\frac{1}{2}-\chi\right)\left(\delta-\frac{h}{2}\right)^2\left[\frac{3(r_i+r_j)}{2}+2\delta+\frac{h}{2}-\frac{3(r_j-r_i)^2}{2(h+r_i+r_j)}\right],$ \hspace{2.6cm}\cite{lozsan:2006}}\\
&\raisebox{-1.0ex}{\hspace{9cm} $\delta < h < 2\delta$.}\\
&\raisebox{-3.0ex}{$V_\mathrm{st}=\frac{k_BT}{V_\mathrm{w}}\left(\frac{1}{2}-\chi\right)\left[\left(\bar{\phi}_j\right)^2\left(\frac{v_a^2}{v_c}-v_a\right)+\left(\bar{\phi}_i\right)^2\left(\frac{v_b^2}{v_c}-v_b\right)+\,2\bar{\phi}_i\bar{\phi}_j\left(\frac{v_av_b}{v_c}\right)\right],$ \hspace{1.7cm}\cite{lozsan:2006}}\\
&\raisebox{-1.0ex}{\hspace{9cm} $0 < h < \delta$.}\\
&\raisebox{-2.0ex}{$V_\mathrm{dil}=\frac{\pi\gamma r_\mathrm{f}^4}{2r_a^2}$. \hspace{10.25cm} \cite{danov:1993a}}\\
\\
&\raisebox{-2.0ex}{$V_\mathrm{bend}=-\frac{2\pi B_0r_\mathrm{f}^2}{r_a}, \, (r_\mathrm{f}/r_a)^2\ll 1$. \hspace{7.43cm} \cite{ivanov:1999}}\\
\raisebox{1.0ex}{}\\
\hline\hline
\end{tabular}
\label{potentials}
\end{table*}

\begin{table*}[th!]
\caption{Tensors used in the simulations. Here $r_i$ is the radius of the small droplet, $r_j$ is the radius of the large drop, $r_*=\frac{2r_ir_j}{r_i+r_j}$, $h$ and $r_\mathrm{f}$ are the thickness and radius of the film, $D_0=\frac{k_BT}{6\pi\eta r_i}$, $\eta$ is the dynamic viscosity of the continuous phase and $\epsilon_\mathrm{S}$ have values between $0.001$ and $1$.}
\centering
\begin{tabular}{l|l}
\hline\hline
Geometry & Tensor\\
\hline\hline
\raisebox{-2.0ex}
{Stokes Immobile: A sphere in an}

&\raisebox{-2.0ex}{$D_\mathrm{St}=D_0$.}\\ 
unbounded liquid \cite{basheva:1999}.\\
\raisebox{-2.0ex}
{Taylor Immobile: Two spheres of radii}
&\raisebox{-2.0ex}{$D_\mathrm{Ta}=4D_0\displaystyle{\frac{r_i}{r_*^2}}h$.}\\
$r_i$,$r_j$, and $h\ll r_i,r_j$ \cite{gurkov:2002}.\\
\raisebox{-2.0ex}
{Two deformed drops \cite{danov:1993b}.}
&\raisebox{-2.0ex}{$D_\mathrm{Dd}=\displaystyle{\frac{4h}{r_i}}\left(1+\displaystyle{\frac{r_\mathrm{f}^2}{r_i h}}+\displaystyle{\frac{\epsilon_\mathrm{S} r_\mathrm{f}^4}{r_i^2 h^2}}\right)^{-1}D_0$.}
\raisebox{-0.5ex}{}\\
\hline\hline
\end{tabular}
\label{diffusion}
\end{table*}

The total potential of interaction between a non-deformable drop and the interface, is assumed to be composed of three contributions: van der Waals, electrostatic, and steric.  In the case of deformable drops, two new potentials of interaction appear during the evolution of the film (Region II). They take into account: (a) the surface deformation energy (extensional or dilational energy) coming from the increase of interfacial area as the spherical drops turn into truncated spheres, and (b) the bending elasticity of the surfactant monolayer adhered to the interface of the drops \cite{ivanov:1999}. These two potentials change with the interparticle distance during the formation of the film (Region II), but reach a constant value after a maximum film radius has been attained. Hence, they do not contribute to the value of the force within Region III.

In order to simulate the movement of the drops, three diffusion tensors were used (Table \ref{diffusion}). In the case of non-deformable drops the expressions of Stokes and Taylor were implemented. In the case of deformable drops, the expression of Danov {\it et al.} \cite{danov:1993b} for deformable drops was used. In the last case, the value of the parameter $\epsilon_s$ was fixed to 1.0, in order to simulate the behavior of tangentially immobile interfaces.

\section{Computational Details}
\label{computational}

The parameters employed in the simulations are shown in Table \ref{parameters}. They correspond to a soybean in water emulsion stabilized with BSA. The protein concentration, pH and ionic strength of the aqueous solution correspond to the the experimental measurements of Basheva {\it et al.} ($4\times10^{-4}$ wt$\%$ protein, pH=$6.4$, ionic strength $0.15\,$M \cite{basheva:1999}). The Hamaker constant was approximated using the expression of Lifshitz for the case of two identical slabs of oil separated by water \cite{israelachvili:1998}. In the absence of more accurate data, the refractive index, dielectric permittivity, and the main absorption frequency in the UV of olive oil were used for this purpose \cite{drummond:1996,olive}. The electric charge of the BSA molecule was calculated by reproducing the value of the $\zeta$ potential ($\zeta$=$-47$ mV) of a $0.5\,\mu$m soybean particle covered by BSA (Kong {\it et al.} \cite{kong:2003}).  The interfacial area of the protein $\Gamma_\mathrm{max}^{-1}$ was also calculated from the data reported on Ref. \cite{kong:2003}. The width of the protein layer was obtained from Ref. \cite{freeman:2004} (see also \cite{jeyachandran:2009,graham:1979}).

\begin{figure}[th!]
\centering
\includegraphics[scale=0.30]{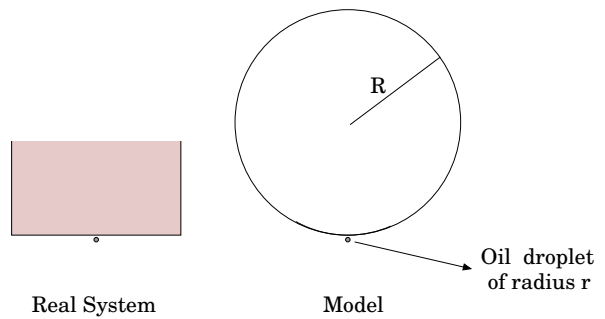}
\caption{Model of the drop/interface system employed in the simulations. Here, $r$ is the radius of the small droplet, and $R$ the radius of the large drop resembling the interface. }
\label{model}
\end{figure}

Following the methodology employed in Ref. \cite{rojas:2010a}, the flat oil/water (O/W) interface was represented by a very large drop of oil fixed in the space (Fig. \ref{model}). A preliminary set of calculations was necessary in order to establish the radius of the large drop and the time step of the simulations. For that purpose, the Flory-Huggins solvency parameter $\chi$ was temporarily set to $0.4$ (for globular proteins it usually varies in the range $0.3$-$0.5$ \cite{tirado:2003}). Radii of $500\,\mu$m, $5000\,\mu$m, and $10,000\,\mu$m were tested for the large drop, and scaled time steps ($\Delta t^*=\Delta t D_0/r_i^2$) of $0.005$, $0.01$, $0.05$, $0.1$ and $1.0$ were studied. The volume fraction of protein around the drops ($\phi$) was used as an input of the simulation ($0.0056 < \bar{\phi}_i < 0.0670$) for several initial distances of separation between the interface (large drop) and the approaching droplets  ($30\,\mathrm{nm} < d < 500\,\mu\mathrm{m}$). The last two variables were systematically changed until the order of magnitude of the experimental coalescence time was reproduced for drops smaller than $r_i < 15\,\mu\mathrm{m}$. It was confirmed \cite{rojas:2010a} that the repulsive potential barriers between the drops had to decrease with the radii of the drops in order to obtain a set of coalescence times that diminish as a function of the particle radius. 

\begin{table}[th!]
\caption{Parameters of the simulations. The molecular properties correspond to one BSA protein.}
\centering
\begin{tabular}{cc}
\hline\hline
\raisebox{-2.0ex}
{Hamaker constant \cite{israelachvili:1998}}
&\raisebox{-2.0ex}{$2.72\times 10^{-21}$ J}\\
\raisebox{-1.0ex}
{Ionic strength \cite{basheva:1999}}              &\raisebox{-1.0ex}{$0.15$ mol/l}\\
\raisebox{-1.0ex}
{Surf. concentration \cite{basheva:1999}}    &\raisebox{-1.0ex}{6.04$\times$10$^{-8}$ mol/l}\\
\raisebox{-1.0ex}
{Molecular mass \cite{czeslik:2004}}               &\raisebox{-1.0ex}{$66.267$ kDa}\\
\raisebox{-1.0ex}
{Molar volume \cite{czeslik:2004}}               &\raisebox{-1.0ex}{$0.048574$ m$^3$/mol}\\
\raisebox{-1.0ex}
{Electric charge}                          &\raisebox{-1.0ex}{$-6.135e$}\\
\raisebox{-1.0ex}
{Area}                                     &\raisebox{-1.0ex}{$11007.81$ \AA$^2$}\\
\raisebox{-1.0ex}
{$\delta$ \cite{freeman:2004}}                                   &\raisebox{-1.0ex}{$1.1\,$nm}\\
\raisebox{-1.0ex}
{$\chi$}                                   &\raisebox{-1.0ex}{$0.49993$}\\
\raisebox{-1.0ex}
{$\gamma$ \cite{basheva:1999}}                                   &\raisebox{-1.0ex}{$15$ mN/m}\\
\raisebox{-0.1ex}{}                        &\raisebox{-0.1ex}{}\\
\hline\hline
\end{tabular}
\label{parameters}
\end{table}

Once the values of $d$ ($d=35\,\mu$m), $\Delta t^*$ (see Table \ref{step}), and $r_j$ ($r_j=5000\,\mu$m) were fixed, the program was allowed to re-calculate $\bar{\phi}_i$ using the following relation:

\begin{equation}
\label{phi}
\bar{\phi}_i=\frac{3r_i^2\Gamma M_\mathrm{p}}{\rho_\mathrm{p}N_A\left[(r_i+\delta)^3-r_i^3\right]},
\end{equation}
where $\rho_\mathrm{p}$ is the density of the protein, $M_\mathrm{p}$ its molecular weight, and $\delta$ the width of the protein layer around the drops. The new values of $\bar{\phi}_i$ predicted by the Eq. (\ref{phi}) were considerably larger than the ones formerly tested (for example, $\bar{\phi}_i=0.662$ for $r_i=6.9\,\mu$m). Hence, the value of $\chi$ was now changed until the order of magnitude of the coalescence time coincided again with the one of the experiments. From this procedure, a value of $\chi =0.49993$ was obtained.

Finally, the values of $\Delta t, r_j, \bar{\phi}_i$, and $\chi$ optimized for non-deformable particles were then used without further modification for the calculations of deformable droplets. However, when Eqs. (\ref{hini_program}) and (\ref{rmax}) were employed to evaluate $h_\mathrm{ini}$ and $r_\mathrm{fmax}$, deformable drops required values of $d$ of the order of nanometers in order to reproduce the order of magnitude of the experimental coalescence time. Instead, non-deformable drops needed initial separations of a few microns.

\begin{table}[htbp]
\caption{Time step of the simulations.}
\bigskip
\begin{tabular}{c|c|c|c}
\hline\hline
 & r ($\mu$m) & $\Delta t^*$ & $\Delta t$ (s)\\
\hline
                  & $1-7.5$    & $1$ & $2.0\times 10^{-6}-8.6\times 10^{-4}$\\
Spherical  & $10-20$   & $10^{-1}$ & $2.0\times 10^{-5}-1.6\times 10^{-4}$\\
                  & $25-40$   & $10^{-2}$ & $3.2\times 10^{-6}-1.3\times 10^{-5}$\\
                  & $80-200$ & $10^{-3}$ & $1.0\times 10^{-6}-1.6\times 10^{-3}$\\
\hline\hline
                       & $25-90$        &  $10^{-2}$ & $3.2\times 10^{-6}-1.5\times 10^{-4}$\\

Deformable & $100-375$     & $10^{-3}$  & $2.0\times 10^{-6}-1.1\times 10^{-4}$\\
                       & $400-1000$ & $10^{-5}$  & $1.3\times 10^{-8}-2.0\times 10^{-7}$\\
\hline\hline
\end{tabular}
\label{step}
\end{table}

\begin{figure}[th!]
\vspace{0.5cm}
\centering
\includegraphics[scale=0.32]{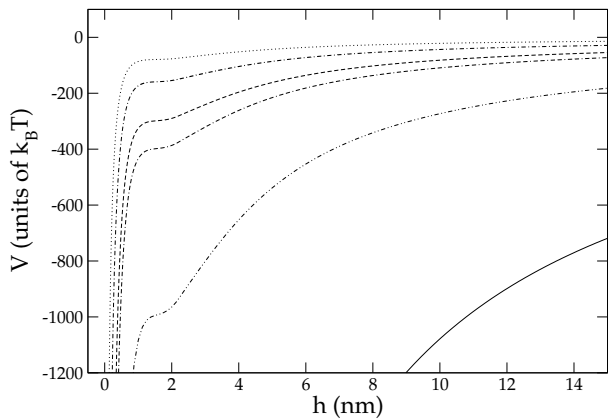}
\caption{Total potential of interaction between a large spherical drop of $5000\,\mu$m  and a small spherical droplet of micron-size. 
Dotted line: $r=2\,\mu$m;
dot-dashed line: $r=4\,\mu$m;
dashed line: $r=7.5\,\mu$m;
dot double-dashed line: $r=10\,\mu$m;
dash double-dotted line: $r=25\,\mu$m;
solid line: $r=100\,\mu$m.}
\label{potsph}
\end{figure}

\begin{figure}[th!]
\centering
\includegraphics[scale=0.32]{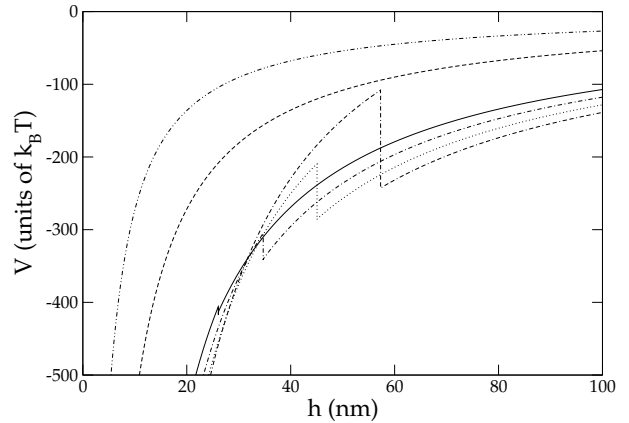}
\caption{Total potential of interaction between a large deformable drop of $5000\,\mu$m  and a small deformable drop of intermediate size ($h_\mathrm{ini}$ and $r_\mathrm{fmax}$ were calculated with Eqs. (\ref{hini_bouyancy}) and (\ref{fr}), respectively). 
Dash double-dotted line: $r=25\,\mu$m;
dashed line: $r=50\,\mu$m;
solid line: $r=100\,\mu$m;
dot-dashed line: $r=110\,\mu$m;
dotted line: $r=120\,\mu$m;
dot double-dashed line: $r=130\,\mu$m.}
\label{def_increase}
\end{figure}

Figures \ref{potsph} and \ref{def_increase} show the potentials of interaction obtained for non-deformable and deformable drops, respectively. The potential of deformable drops shown in Fig. \ref{def_increase} was calculated by using Eqs. (\ref{hini_bouyancy}) and (\ref{fr}) for $h_\mathrm{ini}$ and $r_\mathrm{fmax}$. Notice that the potentials shown in these figures are plotted as a function of $h$ to highlight the location of the repulsive barrier (whenever it occurs). As a result, the width of Region II cannot be appreciated. In the potentials shown in Fig. \ref{def_increase}, the tiny range of distances corresponding to the abrupt jump in potential, marks the growth of the film radius (Region II).  

In the case of non-deformable particles, the total potential of interaction is always attractive, increasing in absolute magnitude as a function of the particle radius. Instead, the potential of deformable droplets shows a very peculiar behavior. For a drop with radius $r_i<100\,\mu$m, it decreases with the increase of the radius. However, for $r_i >100\,\mu$m a repulsive barrier develops. This barrier increases with the particle radius. 

A close look at the partial contributions of the potential of deformable drops indicates that the repulsive barriers illustrated in Fig. \ref{def_increase} are caused by the extensional contribution. At small radii, all the potential contributions occur at a similar distance of separation.  Within a few nanometers, the repulsive contributions decrease in the following order: electrostatic $>$ steric $>$ bending $>$ extensional. However the van der Waals potential prevails for $r_i <100\,\mu$m and the total potential of interaction is attractive at all distances. As the size of the drops increases, $h_\mathrm{ini}$ also grows. Thus, Region II progressively moves toward longer interparticle distances. The range of action of these potentials enlarges (Region III) but the force is exerted within Region II. Thus, the range of action of the extensional and the bending force separates considerably from the one of the electrostatic and the steric contributions. Moreover, the extensional contribution gradually surpasses the bending contribution. At $r_i = 100\,\mu$m, the electrostatic potential shows a peak around 9 nm, while the extensional contribution reaches its maximum around 26 nm. For $r_i = 200\,\mu$m the extensional barrier already occurs at $h=208\,\mathrm{nm}$. Since the absolute magnitude of the van der Waals potential increases with both $r_i$ and $r_f$, the electrostatic and the steric repulsions are always suppressed, and the total potential is attractive at short separations. Yet, the extensional and bending repulsions grow within Region II, exceeding the van der Waals attraction at longer separations (see Fig. \ref{def_increase}). From that point on, a repulsive barrier can be observed in the total potential of interaction.

The present simulations consist in the evaluation of the average coalescence time between a set of drops of different radii and the interface (see Fig. \ref{model}). The average coalescence times were computed from 300 random walks. For preliminary simulations a selected number of particle radii were explored. For the final calculations an extensive number of radii were computed. From $1\,\mu$m to $4.75\,\mu$m the radius was changed using increments of $0.25\,\mu$m. From $10\,\mu$m to $50\,\mu$m increments of $5\,\mu$m were utilized. From $50\,\mu$m to $100\,\mu$m, $10\,\mu$m-increments were employed. Finally, for the range $125\,\mu$m to $1000\,\mu$m, increments of $25\,\mu$m were used. For each particle radius the average coalescence time was first evaluated for spherical particles, and then repeated for deformable drops.

The experimental data in the range $1.48\,\mu$m to $234\,\mu$m was kindly provided by Dr. T. D. Gurkov. The data corresponding to larger particle radii, was obtained directly from Fig. 2 of Ref. \cite{basheva:1999} using the program {\it Engauge Digitizer}.

\section{Results and discussion}
\label{results}

\subsection{Average Value of the Lifetime of the Drops as a Function of the Particle Radius}
\label{average}

The apparatus used in the experiments published by Basheva {\it et al.} \cite{basheva:1999} essentially consisted on a glass vessel with a planar oil/water, and a thin glass capillary situated underneath, $0.5\,$cm below the surface. Different procedures were applied for the making of small drops (micrometer size up to $r_i \leq 100 \mu$m), medium size drops ($100\,\mu$m $< r_i < 500\,\mu$m) and large drops ($r_i>500\,\mu$m). Micrometer size drops were produced by preparing an emulsion with a rotating blade homogenizer. Thus, they were introduced in the experimental cell by means of a glass capillary and a syringe. Medium size droplets were formed squeezing out a drop of oil from the capillary, and then, sucking back the oil in a sudden and fast manner. This produced a turbulent flow of oil and water that enter the capillary, favoring the formation of an emulsion in situ. That emulsion was pushed out subsequently. For large drops a single drop of oil was created at the tip of the capillary and then it was blown out by a certain volume of solution with the help of an additional glass tube. A CCD camera mounted on a microscope was used for recording the process. A timer allowed determining the droplet lifetime, measured from the moment in which the oil droplet begins to move slowly until it coalesces with the bulk oil phase. 

Unlike the above definition given by Basheva {\it et al.} \cite{basheva:1999}, we will use the terms ``medium-size'' or ``intermediate-size'' to refer to those drops in the range  ($10\,\mu\mathrm{m}< r_i < 500\,\mu\mathrm{m}$). 

\begin{figure}[th!]
\vspace{0.5cm}
\centering
\includegraphics[scale=0.32]{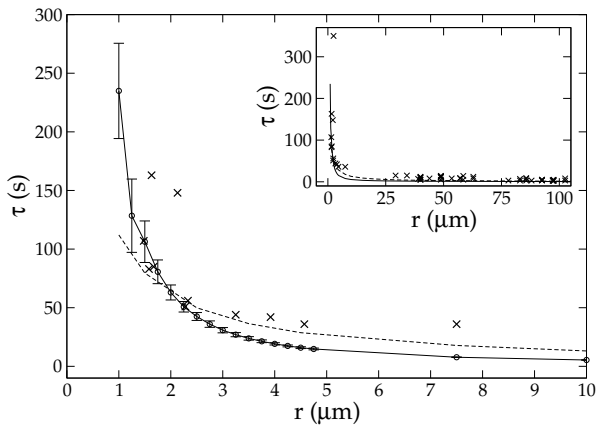}
\caption{Average coalescence time vs. droplet radius for spherical non-deformable drops.  
Crosses: experimental data \cite{basheva:1999};
solid line: constant initial distance of separation between the drops and the interface ($d=35\,\mu$m);
dashed line: variable initial distance of separation between the drops and the interface, $d(r)$. See text for details.}
\label{sphe_d=var}
\end{figure}

Fig. \ref{sphe_d=var} illustrates the results of the calculations for non-deformable micrometer drops which showed the best agreement with the experimental data. The lines in the figure are only a guide to the eye. As previously discussed, the data of micrometric drops was used to adjust the parameters of the steric potential. Notice that the data of Bahseva {\it et al.} \cite{basheva:1999} has a considerable dispersion at all ranges of particle sizes (see below). Due to this fact, it is not possible to reproduce the points corresponding to drops of micrometer size with the same accuracy previously achieved for the experimental data of Dickinson {\it et al.} \cite{dickinson:1988}. The theoretical curve was produced using a fixed distance between the interface and the initial position of the moving drops ($d=35\,\mu$m, solid line in Fig. \ref{sphe_d=var}). It is observed that the lifetime of small drops decreases as a function of the particle size.

The use of a distinct initial distance of approach $d(r_i)$ for drops of different sizes (dashed line), favors a better quantitative agreement between some of the points of the experimental data and the simulations (Fig. \ref{sphe_d=var}). However, it increases the difficulty of the calculations considerably, introducing an undesirable additional variable that needs to be adjusted. This point is illustrated in Fig. \ref{sphe_d=var} using a limited number of simulations. In order to imitate the effect of a variable $d(r_i)$, Eq. (\ref{Stokes-Taylor}) was equalized to the empirical expression of the lifetime of the drops found by Basheva {\it et al.} \cite{basheva:1999}:  $\tau = 0.145/r_i$ \cite{basheva:1999}. The numerical solution of the resulting algebraic equation (using Eq.(\ref{hc}) for $h_\mathrm{crit}$) yielded a value $h_\mathrm{ini}$ for each particle radius. The values of $h_\mathrm{ini}$ changed from $17$ to $1200\,\mu$m  for ($1\,\mu$m $< r_i < 100\,\mu$m). Following, we approximated $d(r_i)$ by $h_\mathrm{ini}$ and ran simulations of non-deformable drops. Observe that $d$ represents the distance below the interface at which the small drops are released, while $h_\mathrm{ini}$ stands for the separation at which the analytic form of the tensor changes from Stokes to Taylor.
The error bars of the theoretical points were approximated by the standard deviation of 300 random walks. In order to optimize the clarity of the figures, the error bars are only shown in Fig. \ref{sphe_d=var}. The uncertainty of the simulations increases with the decrease of the size of the drops, evidencing the effect of the Brownian motion on the movement of the drops. Notice also that the error bars are already small for medium size drops. This indicates that the scattering of the experimental points corresponding to large drops is not caused by their Brownian motion.

As can be appreciated from the inset of Fig. \ref{sphe_d=var}, the curve corresponding to spherical drops also decreases with the particle radius for medium-size drops. Such behavior is essentially the consequence of two factors. First, the attraction between the drops and the interface increases with the size of the drops. Second, the Taylor tensor  (see Eq.  (\ref{sphere-flat}))  is inversely proportional to the square of the particle radius but directly proportional to the buoyancy force. Hence it increases with the radii of the drops, favoring a hyperbolic decrease of the lifetime of small drops as a function of their size. Consequently, the increase of the coalescence time observed for large drops as a function of their radii, cannot be explained assuming non-deformable drops.

\begin{figure}[th!]
\vspace{0.5cm}
\centering
\includegraphics[scale=0.32]{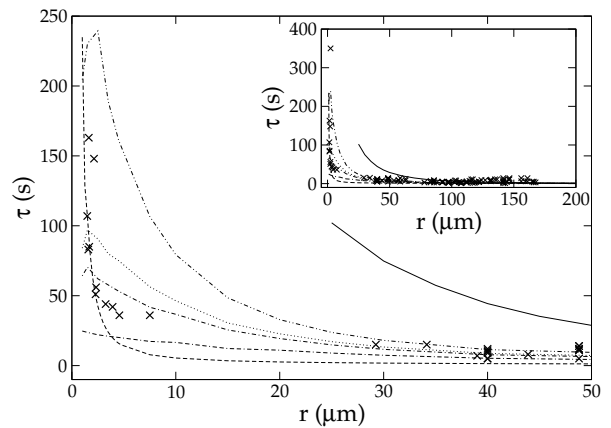}
\caption{Average coalescence time vs droplet radius for deformable droplets.  
Crosses: experimental data \cite{basheva:1999};
solid line: $d=35\,\mu$m;
dashed line: non-deformable drops ($d=35\,\mu$m);
dot double-dashed line: $d=30\,$nm;
dot-dashed line: $d=50\,$nm;
dotted line: $d=60\,$nm; 
dash double-dotted line: $d=100\,$nm.}
\label{all}
\end{figure}

Fig. \ref{all} illustrates similar calculations for deformable drops. In this case, the effect of the initial distance between the drop and the interface is illustrated using distinct values of $d$. In all calculations of deformable droplets shown in this paper the tensor of Danov {\it et al.} \cite{danov:1993a} was used. However, different equations were used to approximate the initial distance of deformation and the value of the maximum film radius. In the simulations of Fig. \ref{all}, Eqs. (\ref{hini_program}) and (\ref{rmax}) were employed to estimate $h_\mathrm{ini}$ and $r_\mathrm{fmax}$. It is observed that initial distances of approach ($d$) of the order of tens of nanometers ($20\,$nm$-100\,$nm) are sufficient to attain the order of magnitude of the experimental data. The change in the value of $d$ with respect to the one of smaller non-deformable drops is due to the dynamics of thinning of the aqueous film between the drop and the interface. The thinning of a plane parallel film is very slow. Hence, a spherical drop needs to diffuse through a longer distance in order to spend a similar period of time before coalescing. 

\begin{figure}[th!]
\vspace{0.5cm}
\centering
\includegraphics[scale=0.32]{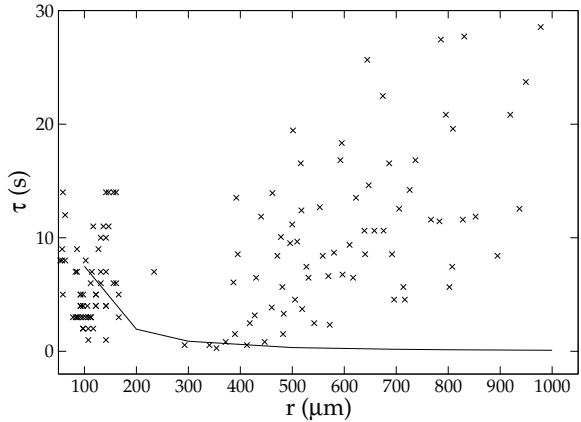}
\caption{Average coalescence time vs droplet radius for deformable drops ($h_\mathrm{ini}$ was calculated using Eq. (\ref{hini_program})).  
Crosses: experimental data \cite{basheva:1999};
solid line: deformable drops ($d=35\,\mu$m.)}
\label{def}
\end{figure}

Unfortunately, the region between $10\,\mu$m $ < r_i <30\,\mu$m does not contain sufficient experimental points to validate which type of drop is more likely to occur. For drops of intermediate sizes the number of experimental points that can be justified using simulations of deformable drops is considerably larger. However, the curves of deformable drops do not overlap smoothly with ones of spherical droplets at small particle radii. Moreover, they show an abnormal decrease close to the hyperbolic rise of the coalescence time, exhibited by the experimental points and the simulations of non-deformable particles.

\begin{figure}[th!]
\centering
\includegraphics[scale=0.32]{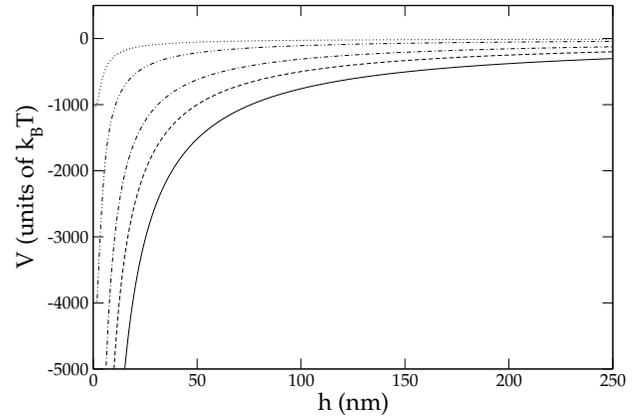}
\caption{Total potential of interaction between a large deformable drop of $5000\,\mu$m  and large deformable drop of a few hundred microns ($h_\mathrm{ini}$ was calculated using Eq. (\ref{hini_program})). 
Dotted line: $r=25\,\mu$m;
dash double-dotted line: $r=100\,\mu$m;
dot-dashed line: $r=300\,\mu$m;
dashed line: $r=500\,\mu$m;
solid line: $r=800\,\mu$m.}
\label{def_typehi0}
\end{figure}

Changes in $d$ from $30\,$nm and $100\,$nm improve the agreement between the experimental data and the calculations. Such a variation appears feasible due to the procedure employed for making the drops of micrometric size. An emulsion is first prepared, and their drops are released from a capillary below the interface. In fact, according to Basheva {\it et al.} \cite{basheva:1999} the experimental procedure allows to measure the size of the drops with an accuracy of $0.5\,\mu$m. Therefore, it is unlikely that changes in the velocity of approach corresponding to separation distances ($d$) lower than $500\,$nm can be determined.

\begin{figure}[th!]
\centering
\includegraphics[scale=0.32]{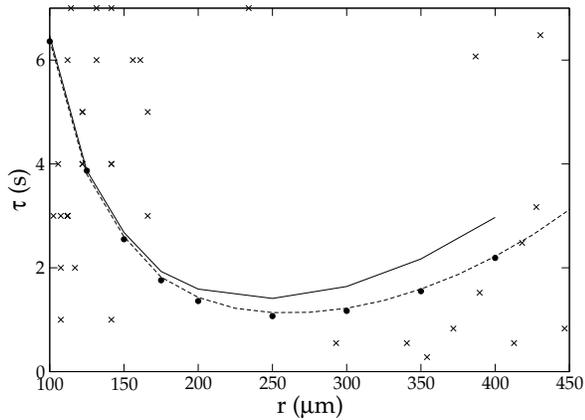}
\caption{Dependence of the lifetime on the potential of interaction between the drops and the interface.
Crosses: experimental data \cite{basheva:1999};
solid line: the movement of the drops is determined by their thermal interaction with the solvent, and the buoyancy force. In these simulations the potential of interaction was completely suppressed, and the tensor of Danov {\it et al.} \cite{danov:1993b} was used, $h_\mathrm{ini}$ was calculated by means of Eq. (\ref{hini_bouyancy}) with $f_r=7/20$;
solid black circles: similar calculations but including the van der Waals potential; 
dashed line: similar simulations in which the total potential of interaction was included.}
\label{onoff}
\end{figure}

When the predictions of the simulations of spherical and deformable droplets are extended to larger particle sizes ($10\,\mu$m to $500\,\mu$m), the curves of $\tau$ vs. $r_i$ monotonically decrease (Figs. \ref{sphe_d=var},  \ref{all} and \ref{def}). In the case of deformable drops (inset of Fig. \ref{all} and Fig. \ref{def}), this is caused by the expressions of $h_\mathrm{ini}$ and $r_\mathrm{fmax}$ used in the simulations. When the radii of the particle increases beyond $100\,\mu$m, the variation of $h_\mathrm{ini} (r_i, \gamma)$ vs. $r_i$ predicted by Eq. (\ref{hini_program}) approach an asymptotic limit of a few tens of nanometers. As a consequence, the increase of $r_\mathrm{fmax}$ as a function of $r_i$ is also dampened, and the potential of interaction decreases with the increase of the particle radius (Fig. \ref{def_typehi0}). However, it is not the increase of the van der Waals attraction which causes the decrease of the coalescence time as a function of the particle radius.  Figure \ref{onoff} shows the results of a limited number of simulations of deformable drops in which no interaction forces were used. For this case, Eqs. (\ref{hini_bouyancy}) and (\ref{fr}) were employed. Neither the extensional nor the bending potentials were included in the simulations. In this case, the lifetime of the drops increases with the size of the drops. Moreover, if the same simulations are run including the van der Waals potential (solid black circles in Fig. \ref{onoff}), an analogous result is obtained. This illustrates the effect of the hydrodynamic tensor on the movement of the drops. It also evidences the influence of the expressions of $h_\mathrm{ini}$ and $r_\mathrm{fmax}$ in the calculations (see below). 

\begin{figure}[th!]
\vspace{0.5cm}
\centering
\includegraphics[scale=0.32]{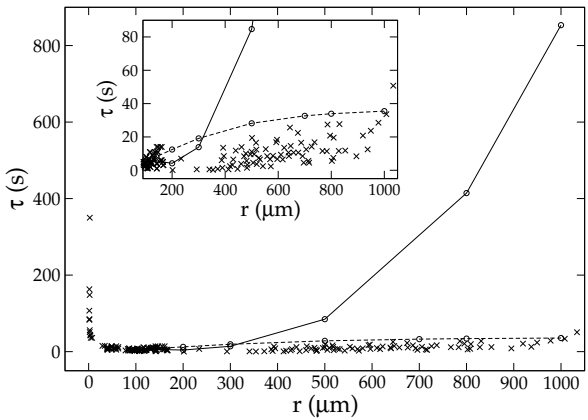}
\caption{Average coalescence time vs droplet radius for deformable drops.  
Crosses: experimental data \cite{basheva:1999};
solid line: $d=35\,\mu$m (Eq.  (\ref{hini_bouyancy}) with $f_h=1$ and Eq. (\ref{fr}) with $f_r=1$ were used as approximations for $h_\mathrm{ini}$ and $r_\mathrm{fmax}$);
dashed line: $d=10\,\mu$m with $r_\mathrm{f}=0.05\, r_i$, and $h_\mathrm{ini}=10^{-5}\,$m}
\label{def_f=1}
\end{figure}

Fig. \ref{def_f=1} shows the results of similar simulations in which the total potential of interaction was included. This potential increases with the particle radius as a consequence of the extensional contribution (Fig. \ref{def_increase}). As in the previous calculations, Eqs.  (\ref{hini_bouyancy}) and (\ref{fr}) were used as approximations for $h_\mathrm{ini}$ and $r_\mathrm{fmax}$. The coalescence time increases pronouncedly with the radii of the drops. Surprisingly, the increase of the lifetime of the drops as a function of the particle radius is not caused by the development of the repulsive potential barrier. This can be realized comparing the dashed curve and the black circles illustrated in Fig. \ref{onoff}: the lifetimes predicted by the simulations in the presence or absence of repulsive forces are similar.

Table \ref{forces} shows a comparison between the largest magnitude of the repulsive force (caused by the extensional deformation of the drops) and the value of the buoyancy force. It is observed that in the present case the buoyancy force subdues the effect of the interaction forces. Consequently, our results coincide with the previous predictions of Ivanov and Kralchevsky \cite{ivanov:1997} for the case of zero disjoining pressure: the lifetime of large drops increases as a function of their radii (see Fig. 3 in Ref. \cite{ivanov:1997}). 

\begin{table}[th!]
\caption{Intermolecular and Bouyancy force.}
 \begin{tabular}{c|c|c}
\hline\hline
 $r$ ($\mu$m) & $F_\mathrm{max}=-\left(\frac{\partial V}{\partial r_{ij}}\right)_\mathrm{max} (N)$ & $|F|=\frac{4}{3}\pi r_i^3\Delta\rho g   
 \,\,(N)$\\
\hline\hline 
$25$      & $-1.35 \times 10^{-10}$ & $4.95 \times 10^{-11}$\\
$50$      & $-2.25 \times 10^{-12}$ & $3.96 \times 10^{-10}$\\
$100$    & $ 1.21 \times 10^{-11}$ & $3.16 \times 10^{-9}$\\
$250$    & $ 2.24 \times 10^{-9}$   & $4.95 \times 10^{-8}$\\
$500$    & $ 1.71 \times 10^{-8}$    & $3.96 \times 10^{-7}$\\
$750$    & $ 5.61 \times 10^{-8}$    & $1.33 \times 10^{-6}$\\
$1000$ & $1.30 \times 10^{-7}$     & $3.16 \times 10^{-6}$\\
\hline\hline
 \end{tabular}
\label{forces}
\end{table}

It is important to realize that the extensional potential only acts within Region II. The width of this region depends on the size of the drops, changing between $30\,$nm and $1.92\,\mu$m in the range $250\,\mu$m $< r_i < 1000\,\mu$m. Therefore a limited number of additional simulations were run decreasing the time step in one and two orders of magnitude, to improve the sampling of the repulsive barrier. However, similar lifetimes were obtained. This fact validated the magnitude of the time steps used in the simulations. Furthermore, it verifies that the extensional contribution of the potential was unable to overcome the effect of the buoyancy force.   

We also tested several alternative expressions for $h_\mathrm{ini}$ and $r_\mathrm{fmax}$. The dashed line of Fig. \ref{def_f=1} illustrates the result of using  $h_\mathrm{ini}$ as an input of the simulations ($h_\mathrm{ini}=10^{-5}\,$m) along with a slight modification of Eq. (\ref{rmax}): $r_\mathrm{fmax}=f_r \sqrt{r_ih_\mathrm{ini}}$ ($f_r=0.055$).  Some of these expressions favor an increase of $\tau$ as a function of $r_i$ but produce a convex curve contrary to the experimental results.  It is possible that equations of the form: $h_\mathrm{ini} \propto (r_i)^n$, and $r_\mathrm{fmax} \propto (r_i)^m$  might be valuable in the future, but we did not make an extensive study of these expressions. In any event, it is evident that the predictions of the simulations for large drops depend sensibly on the analytic form of $h_\mathrm{ini}$ and $r_\mathrm{fmax}$.

\begin{figure}[th!]
\centering
\includegraphics[scale=0.32]{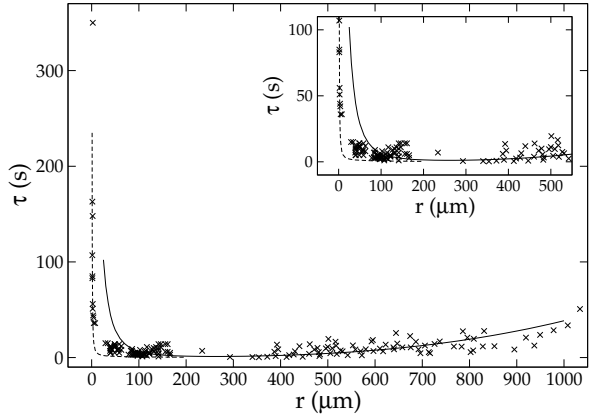}
\caption{Average coalescence time vs droplet radius for spherical and deformable drops initially separated by a distance of $d=35\,\mu$m. 
Crosses: experimental data \cite{basheva:1999};
dashed line: simulation with spherical drops;
solid line: simulation with deformable drops.}
\label{final_d=35}
\end{figure}

In order to reproduce the experimental value of $\tau$ for $r_i=1000\,\mu$m, the fraction $f_r$ was systematically changed while keeping $f_h=1$ in Eq. (\ref{hini_bouyancy}). Table \ref{f} shows the coalescence times obtained for different values of $f_r$. The results of this table suggest that a value of $f_r =7/20$ might be adequate to reproduce the experimental data of large drops. Fig. \ref{final_d=35} shows the results of the simulations for the complete range of $r_i$ using $f_r=7/20$ and a constant value of $d=35\,\mu$m. This is the value of $d$ previously employed in the calculation of spherical drops of micrometric size. For particle radii in the range $1-10\,\mu$m we already showed that the simulations of non-deformable drops reasonably reproduce the experimental data. It is now observed that in the range $100\,\mu$m $< r_i < 1000\,\mu$m the experimental data is also recreated using deformable drops if Eq. (\ref{hini_bouyancy}) is used to approximate the initial distance of deformation and Eq. (\ref{fr}) is employed to calculate $r_\mathrm{fmax}$ with the factor $f_r=7/20$. 

Despite the above results the extension of the simulations of deformable droplets to the small range of particle sizes, fails to reproduce the experimental data (Fig. \ref{final_d=35}). Since the tensor of Danov {\it et al.} tends to the expression of Taylor for small particle radius, $\tau$ increases when $r_i$ decreases. However, this augment occurs at considerably larger particle radii than those experimentally observed. Moreover, the expression of $h_\mathrm{ini}$ calculated by means of Eq. (\ref{hini_bouyancy}), becomes smaller than $h_\mathrm{crit}$ for $r_i < 25\,\mu$m. This unphysical result is partially due to the use of Eq. (19) to approximate $h_\mathrm{crit}$. This equation was deduced assuming that only van der Waals forces occur in the system. In any event, it is not possible to do simulations for $r_i < 25\,\mu$m using Eq. (\ref{hini_bouyancy}) to approximate $h_\mathrm{ini}$. 

\begin{figure}[th!]
\vspace{0.6cm}
\centering
\includegraphics[scale=0.32]{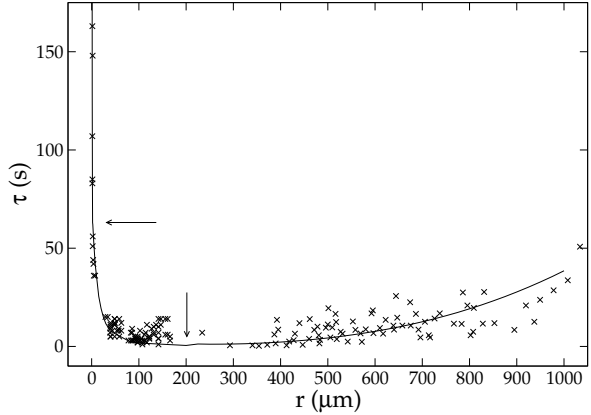}
\caption{Average coalescence time vs. droplet radius for spherical and deformable drops. 
Crosses: experimental data \cite{basheva:1999};
solid line: simulations of spherical and deformable drops. The picture illustrates the results of three different methodologies of calculation for three distinct regions of particle sizes. The arrows indicate the overlap between  these regions.}
\label{final_d=op}
\end{figure}

In general, as the particle radius decreases, the effect of the buoyancy force is less relevant. The influence of the buoyancy force on the lifetime of large drops is decisive despite the considerable magnitude of the extensional potential. At the other extreme of particle sizes, the experimental data of micron-size drops $r_i < 10\,\mu$m can only be reproduced using attractive potentials and non-deformable drops. For these drops the magnitude of the attractive potential increases with the size, reinforcing the effect of the hydrodynamic tensors (Stokes and Taylor). Moreover, the effect of the Brownian motion is significant as evidenced by the error bars of the simulations. For medium-size particles, the errors bars are very small and the outcome of the simulations depends on the interplay between the buoyancy force and the potential of interaction. However, as the particle radius decreases, the deformation of the drop also diminishes. Consequently, the repulsive extensional potential decreases significantly along with the buoyancy force. Hence, the effect of the buoyancy force still predominates while the deformation of the drops decreases and the hydrodynamic tensor approaches the form of the Taylor tensor. 

An empirical fit of the computed lifetime of droplets for $300 \leq r_i \leq 1000\,\mu$m, leads to:

\begin{equation}
\label{tau_1}
\tau = 3.3494 \times 10^{-8} r_i^{3.0138}, \quad (r^2 = 0.9985).
\end{equation}

It is remarkable that the value of the exponent in Eq. (\ref{tau_1}) is fairly close to $25/7 = 3.57$ suggested by Basheva {\it et al.} \cite{basheva:1999}. The theoretical estimation results from substituting the critical thickness of rupture suggested by Vrij \cite{vrij:1966, vrij:1964, vrij:1968}:

\begin{equation}
 \label{hc}
h_\mathrm{crit}=0.268\left(\frac{36\pi^3 A_H^2 r_\mathrm{f}^4}{6.5F\sigma}\right)^{1/7}.
\end{equation}
into Eq. (\ref{tau}):

\begin{equation}
 \label{tau_2}
\tau=4.088\,\eta\sigma^{-8/7}A_H^{-4/7}\left(\Delta\rho g\right)^{5/7} r_i^{25/7}.
\end{equation}

\begin{table}[th!]
\caption{Choice of the parameter $f_r$.}
\centering
\begin{tabular}{c|c|c|c|c}
\hline\hline
\raisebox{-2.0ex} {$f_r$}&\raisebox{-2.0ex} {\hspace{0.2cm} $\displaystyle{\frac{1}{2}}$\hspace{0.5cm}}&\raisebox{-2.0ex} {\hspace{0.5cm} $\displaystyle{\frac{2}{5}}$\hspace{0.5cm} }&\raisebox{-2.0ex} {\hspace{0.5cm} $\displaystyle{\frac{7}{20}}$\hspace{0.5cm} }&\raisebox{-2.0ex} {\hspace{0.5cm} $\displaystyle{\frac{3}{10}}$\hspace{0.5cm} }  \\
 \raisebox{-0.1ex} {}&\raisebox{-0.1ex} {}&\raisebox{-0.1ex} {}&\raisebox{-0.1ex} {}&\raisebox{-0.1ex} {}\\
\hline
 \raisebox{-2.0ex} {$\tau (s)$}&\raisebox{-2.0ex} {\hspace{0.3cm}$110.37$\hspace{0.3cm}}&\raisebox{-2.0ex} {\hspace{0.3cm}$57.10$\hspace{0.3cm}}&\raisebox{-2.0ex} {\hspace{0.3cm}$38.49$\hspace{0.3cm}}&\raisebox{-2.0ex} {\hspace{0.3cm}$24.42$\hspace{0.3cm}}\\

 \raisebox{-1.0ex} {}&\raisebox{-1.0ex} {}&\raisebox{-1.0ex} {}&\raisebox{-1.0ex} {}&\raisebox{-1.0ex} {}\\
\hline\hline
\end{tabular}
\label{f}
\end{table}

As it was explained at the beginning of this section, the experimental procedure employed for the production of drops of different sizes was distinct for drops of micrometric, medium size, and large size. Fig. \ref{final_d=op} illustrates the lifetime of the simulations which render the closest agreement with the experimental data.  For $r_i < 2\,\mu$m spherical drops were calculated using $d=35\,\mu$m. In the range $2.5 \,\mu$m $< r_i < 200\,\mu$m deformable drops were simulated using Eq. (\ref{hini_program}) for $h_\mathrm{ini}$ and $d=50\,$nm. Finally, for the range $225\,\mu$m $< r_i < 1000\,\mu$m the experimental data was reproduced considering deformable drops with an initial distance of approach of  $d=35\,\mu$m ($f_r=7/20$).  It is evident that Eq. (\ref{hini_program}) is more reliable than  Eq. (\ref{hini_bouyancy}) in order to approximate the value of $h_\mathrm{ini}$ in the range $15 \,\mu$m $< r_i <100\,\mu$m. Since the difference between the experimental data and the theoretical prediction is basically caused by the method of evaluation of $h_\mathrm{ini}$ and $r_\mathrm{fmax}$, it is also reasonable to conclude that the drops behave as deformable particles over the intermediate range of particle sizes. 

\begin{figure}[th!]
\vspace{0.6cm}
\centering
\includegraphics[scale=0.32]{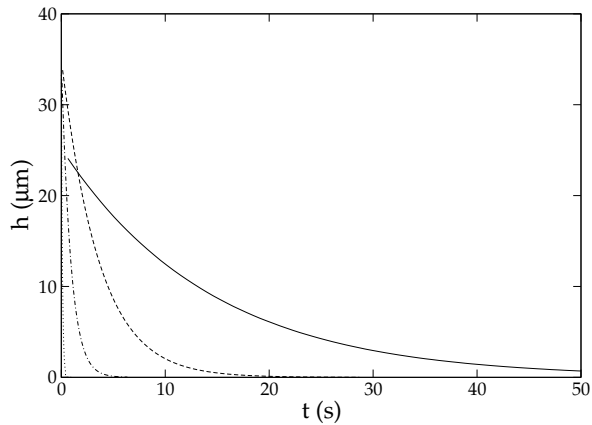}
\caption{Variation of film thickness with time. 
Solid line: $r=25\,\mu$m;
dashed line: $r=50\,\mu$m;
dot-dashed line: $r=100\,\mu$m;
dotted line: $r=300\,\mu$m.}
\label{hvst}
\end{figure}

For completeness, the change of $h$ vs. $t$ between the drops and the interface was calculated for different particle radii (Fig. \ref{hvst}). For drops larger than $25\,\mu$m, the thinning of the film is monotonous and the effect of the Brownian movement small. Using the data of $h$ vs. $t$, the velocity of thinning, $-\d h/\d t$ was evaluated numerically at $t =1\,\mathrm{s}$ for a selected set of particle radius belonging to the region of large drops ($r_i \geq 500\,\mu$m).

Several models are available in the literature for the calculation of rate of drainage $-\d h/\d t$ between a drop and the interface in the absence of surfactant molecules  \cite{mackay:1963}.  They reproduce the cases in which: (a) the sphere is non-deformable but the interface deforms; (b) the sphere and the interface form a plane parallel film; (c) the drop and the interface deform forming a curved film; and (d) the drop forms a dimple and the interface does not deform \cite{mackay:1963}. The first three models (a)$-$(c) predict that:

\begin{equation}
\label{dhdtvsh}
 -\frac{\d h}{\d t} = \mathrm{c }\, h^3,
\end{equation}
with a different coefficient $\mathrm{c}$. Despite their limitations, the referred formalisms assume the occurrence of tangentially immobile interfaces as it happens in the presence of surfactants molecules. The dependence of $-\d h/\d t$ on $h$ comes from the use of the Stephan-Reynods equation \cite{reynolds:1881}.
Most experiments that measure $h_\mathrm{crit}$, and $-\d h/\d t$ usually involve microscopic distances between film radius of macroscopic size (see Fig. 2 in Ref. \cite{velikov:1997}). Such experiments indicate that there is a simple proportionality between $-\d h/\d t$ and $h$ for a wide range of film thickness ($25\,$nm to $100\,$nm) \cite{manev:2005a}. According to our results: 

\begin{equation}
\label{dhdt_our}
 -\frac{\d h}{\d t} = 0.77\, h^{0.91}, \quad r^2=0.9997.
\end{equation}

The exponent of Eq. (\ref{dhdt_our}) appears to be reasonable considering the limitations of the simulations (constant $d$, plane parallel film of uniform thickness, etc.). This proportionality between the velocity of thinning and its width can be partially justified analytically. If the Brownian motion of the large drops is assumed to be negligible, their velocity is only the product of the diffusion tensor and the buoyancy force:

\begin{equation}
\label{v}
 V=D_\mathrm{Danov} \frac{F}{k_BT}.
\end{equation}

Using the tensor of Danov {\it et al.} \cite{danov:1993a} and Eq. (\ref{fr}) to relate the film radius with the particle radius gives:

\begin{equation}
 \label{v_simplify}
 -\frac{\d h}{\d t}=\frac{\bar{c}_1  h^3} {h^2+ \bar{c}_2 h + \bar{c}_3 }.
\end{equation}

The coefficients of Eq. (\ref{v_simplify}) depend on: $\Delta\rho,g,\gamma,f_r,\epsilon_s$ and $\eta$. The relative magnitude of the coefficients determine the power dependence of the thinning velocity on $h$. For $r_i=100\,\mu$m: $\bar{c}_1=67.18\,\mathrm{s}^{-1}$, $\bar{c}_2=1.69\times 10^{-17}\,$m,   and $\bar{c}_3=4.11\times 10^{-8}\,$m$^2$. Likewise, for  $r_i=1000\,\mu$m: $\bar{c}_1=671.82\,\mathrm{s}^{-1}$, $\bar{c}_2=1.69\times 10^{-11}\,$m,   and $\bar{c}_3=4.11\times 10^{-6}\,$m$^2$.  It is clear then that for the parameters corresponding to the experiments of Basheva {\it et al.} \cite{basheva:1999}, the last two terms of the denominator of Eq. (\ref{v_simplify}) are negligible and as a result: $-\d h/\d t  \propto h$.

\subsection{Dispersion of Experimental Lifetimes. Possible Role of Interaction Forces and Adsorption Times }
\label{dispersion}

It is clear from Figs. \ref{def} and \ref{final_d=op} that the experimental lifetimes show a considerable scattering at large particle radii. 
Gurkov {\it et al.} \cite{gurkov:2002} had pointed out that this phenomenon is caused by the statistical nature of the coalescence process. The rupture of thin liquid films might occur through the spontaneous growth of fluctuation waves on the two opposing O/W interfaces of the film, or through the formation of holes \cite{nikolova:1999,kashchiev:1980}. In either case the rupture of the film is a stochastic process and must be treated accordingly. Gurkov {\it et al.} \cite{gurkov:2002} studied the variation of the number of drops, $N(t)$, which survive coalescence until time $t$ ($1000\,\mu$m $\leq r_i \leq 3,500\,\mu$m). The relative change ($w = 1/N \times dN/dt$) is related to the probability for ``drop burst per unit time''. Since $w$ is time dependent, it can be expanded in power series:

\begin{equation}
 \label{N}
-\log \frac{N}{N_\mathrm{tot}}= w_0 (t-t_0) + w_1 (t-t_0)^2 + \cdots,
\end{equation}
where $t_0$ is the time when coalescence is first observed in the system. Surprisingly, use of Eq. (\ref{N}) produces a monotonous parabolic curve of $-\log \frac{N}{N_\mathrm{tot}}$ vs. $t-t_0$, where ${N_\mathrm{tot}}$ is the total number of drops studied  (see Fig. 6 in Ref. \cite{gurkov:2002}).

According Ghosh and Juvekar \cite{ghosh:2002}, the scattering of the rest time of the drops is caused by differences in the surface excess of the surfactant corresponding to each drop. Those differences promote distinct magnitudes of the repulsive force between the drop and the interface. That force acts mainly on the ``barrier ring'' of the O/W/O film due to heterogeneity of the film thickness and the outward flow generated by the collision of the drop with the interface. The analytic treatment of the problem leads to a probability distribution of rest times which can be nicely fitted to the experimental data whenever surfactant is present in the system. Systems without added surfactant show a considerable degree of irreproducibility. 

The results presented in the previous sections reinforce the belief that it is the hydrodynamic resistance and not the interaction forces that determine the characteristic variation of the coalescence time of the drops as a function of their radii. It was shown that the deformation of the drops between $100\,\mu$m and $1000\,\mu$m, promotes the development of a significant repulsive barrier which grows as a function of the particle radius. Despite this fact, the buoyancy force is so large that it suppresses the effect of the potentials on the lifetime of the drops. When this happens, the behavior found is necessarily determined by the mathematical form of the hydrodynamic tensor. 

Up to this point the results correspond to the case in which a steady thinning of the intervening liquid between a drop and the interface occurs. Therefore they cannot explain the large scattering of lifetimes shown by the large drops. We explored the possibility that the referred scattering could be caused by the contact of the capillary waves formed at each oil/water interface of the film. Thus, every time that a drop enter the regions of deformation II and III, the program starts to measure the lifetime of the film, $\tau_{ij}$. Coalescence occurs when

\begin{equation}
 \lambda_\mathrm{TOTAL}=(\lambda_i+\lambda_j) \left[ \exp\left(\frac{\tau_{ij}}{\tau_{Vrij}}\right)-1\right],
\end{equation}
is larger than the width of the film. In the equation above

\begin{equation}
 \label{lambda}
\lambda_i(t)=\mathrm{Ran} (t)\, h_\mathrm{crit}.
\end{equation}

Here, $\mathrm{Ran}(t)$ stands for a random number between $-1.0$ and $1.0$. The actual lifetime of a film ($\tau_{ij}$) between drops $i$ and $j$ is compared -at each time step- with the analytical equation of Vrij \cite{vrij:1968} for the fastest increase of surface oscillations:

\begin{equation}
 \label{tauvrij}
\tau_{Vrij}=96\pi^2\gamma\eta h_0^5 A_H^{-2}.
\end{equation}

Eq. (\ref{tauvrij}) was deduced assuming that only van der Waals forces occur between the drops. As a consequence, it is generally small and the methodology does not have an appreciable influence on the coalescence time of drops of micron size. However, the value of $\tau_{Vrij}$ depends sensibly on the thickness of the films, which is large for intermediate and large drops. It is important to remark that in the simulations $\tau_{Vrij}$ is recalculated at each iteration. Therefore, it changes continuously with $h$.

Unfortunately, preliminary calculations indicated that the methodology described was unsuccessful in reproducing the experimental scattering of the lifetimes of the drops when the potentials shown in Figures \ref{potsph} and \ref{def_increase} are used in the simulations. It was already shown that the buoyancy force is greater than the extensional (and the bending) force which occur within region II. Hence, there is no repulsive barrier to prevent the rapid thinning of the film. When the mechanism of capillary waves is implemented, coalescence occurs at an even faster rate, but this rate is reproducible and does not give rise to a considerable scattering of lifetimes.

It is clear that the scattering of lifetimes can only occur if a substantial repulsive force overcomes the inertia of the particles caused by the buoyancy force. This is necessary independently of the mechanism of coalescence proposed. Out of all the potential contributions studied,  it is the steric contribution the one that can be less accurately determined. The analytic form of the potential used in the previous calculations is the one formerly employed in the estimation of the lifetimes of micron-size drops \cite{rojas:2010a}. Thus, it might not be suitable for the simulation of large deformable drops. Hence, a different expression of a steric potential with a harder repulsive barrier was tested. This equation results from the adjustment of the potential proposed by Alexander-De Gennes \cite{alexander:1977,degennes:1987} to the case of truncated spheroids \cite{danov:1993a}:

\begin{eqnarray}
 \label{danov:3.48}
V_\mathrm{st}&=&\pi r_\mathrm{f}^2f(h)+4\pi r_i k_B T\, \Gamma^{3/2} L_g^2 \times\\
\nonumber
&&\left[1.37 h_g-0.21 h_g^{11/4} + 3.20 h_g^{-1/4}-4.36\right],\\
\nonumber
f(h)&=&2k_B T\,\Gamma^{3/2} L_g \left[\frac{4}{5}h_g^{-5/4}+\frac{4}{7}h_g^{7/4}-1.37\right],
\end{eqnarray}
where $h_g=h/(2 L_g)$, $L_g=N_\mathrm{seg}\left(\Gamma l_\mathrm{seg}^5\right)^{1/3}$. $N_\mathrm{seg}$ is the number of segments of the protein ($580$ aminoacids \cite{jeyachandran:2009}), $l_\mathrm{seg}$ is the length of each segment ($\approx 3.0 \times 10^{-10}\,$m) and $\Gamma$ is the number of molecules per unit area ($9.08 \times 10^9$ molecules/m$^2$). The value of $l_\mathrm{seg}$ was approximated by $V_a^{1/3}$, where $V_a$ is the typical volume of an amionoacid residue ($V_a=57\,\AA^3-186\,\AA^3$) \cite{creighton:1984}. 

\begin{figure}[th!]
\vspace{0.6cm}
\centering
\includegraphics[scale=0.33]{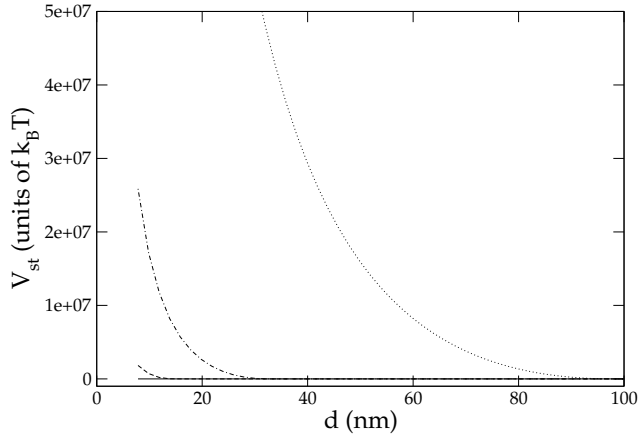}
\caption{Steric potential. 
Solid line: steric potential used of the simulations of the Figs. \ref{sphe_d=var}-\ref{hvst};
dashed line: steric potential of Eq. (\ref{danov:3.48}) with $l_\mathrm{seg}=1.85\AA$;
dot-dashed line: steric potential of Eq. (\ref{danov:3.48})  with $l_\mathrm{seg}=3.00\AA$;
dotted line: steric potential of Eq. (\ref{danov:3.48})  with $l_\mathrm{seg}=5.80\AA$.}
\label{steric}
\end{figure}

It was observed that the lifetime predicted by the calculations depends considerably on the average length of a protein segment. For $l_\mathrm{seg}=5.8 \times 10^{-10}\,$m the total potential becomes repulsive at short distances and the drops do not coalesce even if the mechanism of capillary waves is activated. In order to study the influence of the new potential on the mechanism of capillary waves, an approximate value of $l_\mathrm{seg}$  was used ($l_\mathrm{seg}\ =3.0 \times 10^{-10}\,$m). The value of $l_\mathrm{seg}$ was not adjusted to reproduce the order of magnitude of the lifetime of the drops.  We just substituted the steric potential used in the previous simulations (section \ref{average}) for the one of Alexander-de Gennes (Eq. (\ref{danov:3.48})) to appraise the resulting effect. The steric potential used is illustrated in Fig. \ref{steric}. 

As expected, it was observed that the lifetime of the drops increased with the strength of the steric potential. For drops of $r_i=200\,\mu$m,  $400\,\mu$m, and $600\,\mu$m, the value of $\tau$ increased in $15.8$, $17.2$, and $13.3$ seconds, respectively. However, the augment of the standard deviation was not sufficient to justify the large dispersion of the experimental data. The  deviation increased from $0.0017$ to $0.0094$ for $r_i=200\,\mu$m, from $0.0019$ to $0.020$ for $r_i=400\,\mu$m, and from $0.0039$ to $0.019$ for $r_i=600\,\mu$m.

In Emulsion Stability Simulations, the interaction potentials depend on the degree of surfactant adsorption to the O/W interface. Previous calculations \cite{urbina:2009a, urbina:2004a} confirmed that the coalescence time between two drops is inversely proportional to the surfactant surface excess at the interface of the drops. Due to the characteristics of the experimental set up under consideration, it is very likely that drops with different surface excesses might be produced. It was already shown \cite{rojas:2010a} that the dispersion of the lifetime of small drops is caused by their Brownian movement. However, the lifetime of small drops is more reproducible than the one of large drops. This coincides with the fact that small drops are produced by means of emulsification, while large drops are generated in situ. Thus, the protein concentration of the large drops is expected to grow as a function of time, starting from the moment the drop is formed within the aqueous phase. 

In Ref. \cite{urbina:2004a} we studied the problem of time-dependent adsorption and its influence on the coalescence time. Liggieri {\it et al.} \cite{liggieri:1996} demonstrated that several cases of mixed adsorption kinetics can be reformulated into a diffusion-controlled formalism \cite{ward:1946} if the diffusion constant of the molecules ($D_\mathrm{m}$) is substituted by an ``apparent'' diffusion constant $D_\mathrm{app}$. 

Table \ref{dapp} shows the results of additional simulations in which the effect of a time dependent adsorption was considered. In these calculations, the Alexander-De Gennes potential  \cite{alexander:1977,degennes:1987} for truncated spheroids \cite{danov:1993a} was used (Eq. (\ref{danov:3.48}) with $l_\mathrm{seg}=3.00\AA$). The radius of the small drop was kept fixed at $r_i = 200\,\mu$m, and the value of $D_\mathrm{app}$ was changed in several orders of magnitude. The mechanism of coalescence that includes capillary waves was activated. Using these conditions, the surface excess of proteins at the interface of the drops was evaluated from \cite{ward:1946, rosen:1990, hua:1988, hua:1991}: 

\begin{equation}
 \label{gamma}
\Gamma=2 \left(\frac{D_\mathrm{app}}{\pi}\right)^{1/2} C_p\, t^{1/2},
\end{equation}
where $D_\mathrm{app}$ is the apparent diffusion constant, $C_p$ is the protein concentration, and $t$ is the time elapsed from the moment that the small drop is released. The time required for complete protein coverage of the drop can be calculated from the same equation substituting the surface excess by the inverse of the minimum interfacial area of the protein at the oil/water interface: 

\begin{equation}
 \label{tc}
t_c=\left(\frac{\Gamma_\mathrm{max}}{A_c C_p}\right)^2,
\end{equation}
where $t_c$ is the critical time, $\Gamma_\mathrm{max}=9.08\times 10^{15}$ proteins/m$^2$, and $A_c=2 (D_\mathrm{app}/\pi)^{1/2}$.

Notice that when the surface excess of the protein increases as a function of time, the value of the interfacial tension decreases between $\gamma = 50\, \mathrm{mN/m}$ and $\gamma = 15\,\mathrm{mN/m}$. Hence, $h_\mathrm{ini}$, $r_\mathrm{f}$, and the interaction potentials also change as a function of time. Moreover, the value of $r_\mathrm{f}$ is not fixed within region III, but also changes.

From the results shown in Table \ref{dapp}, the following conclusions can be drawn:

\begin{description}

\item[1.] If the apparent diffusion constant of the protein is high, $D_\mathrm{app} > 2.95 \times 10^{-9}$ m$^2$/s, the repulsive potential between the drop and the interface builds up very quickly. Consequently, the coalescence process is substantially slowed down, favoring a lifetime of the order of $17.5$ seconds.
 
\item[2.] If $D_\mathrm{app} < 1.0 \times 10^{-9}$ m$^2$/s, the coalescence of the drop and the interface occurs very fast, $\tau \approx 1.4\,$s. The protein does not arrives in time to delay the thinning of the film.

\item[3.] Intermediate values of $D_\mathrm{app}$ promote a large dispersion of lifetimes. The drops start to feel the repulsive potential at different film thickness. Hence, their coalescence time changes appreciably. Moreover, the standard deviation of $300$ simulations becomes very large, approaching $-$ and surpassing in some cases $-$ the order of magnitude of the average lifetime. The variation of the deviations indicates that for $2.86 \times 10^{-9}$ m$^2$/s  $ \leq D_\mathrm{app} \leq 2.89 \times 10^{-9}$ m$^2$/s, slightly different paths might lead to considerable differences in the lifetime of the drops. These are caused by different values of $\Gamma(t)$.

\end{description}

\begin{table}[th!]
\caption{Variation of the $\tau$ whit the apparent diffussion constant. $\sigma$ represents the standard deviation. } 
 \begin{tabular}{c|c|c|c}
\hline\hline
  $D_\mathrm{app}$ (m$^2$/s) & $\tau$ (s) & $\sigma$ (s) &$t_c$ (s)\\
\hline\hline
$1.000\times 10^{-12}$ & $1.45$   & $7.50 \times 10^{-4}$ & $4.90 \times 10^{4}$\\
$1.000\times 10^{-10}$ &  $1.44$  & $7.13 \times 10^{-4}$ & $4.90 \times 10^{2}$\\
$1.000\times 10^{-9}$   & $1.43$  & $9.19 \times 10^{-4}$ & $49.02$\\
$2.000\times 10^{-9}$   & $1.56$  & $6.88 \times 10^{-3}$ & $24.51$\\
$2.250\times 10^{-9}$   & $1.65$  & $9.60 \times 10^{-3}$   & $21.79$\\
$2.500\times 10^{-9}$   & $1.81$  & $1.48 \times 10^{-2}$   & $19.61$\\
$2.750\times 10^{-9}$   & $2.15$  & $3.32 \times 10^{-2}$   & $17.83$\\
$2.800\times 10^{-9}$   & $2.30$  & $5.03 \times 10^{-2}$   & $17.51$\\
$2.850\times 10^{-9}$   & $2.58$  & $9.95 \times 10^{-2}$   & $17.20$\\
$2.860\times 10^{-9}$   & $2.94$  & $1.91$                            & $17.14$\\
$2.865\times 10^{-9}$   & $4.18$  & $4.38$                            & $17.11$\\
$2.870\times 10^{-9}$   & $7.02$  & $6.66$                            & $17.08$\\
$2.875\times 10^{-9}$   & $10.17$  & $7.34$                          & $17.05$\\
$2.880\times 10^{-9}$   & $13.29$  & $6.64$                          & $17.02$\\
$2.885\times 10^{-9}$   & $14.95$  & $5.55$                          & $16.99$\\
$2.890\times 10^{-9}$   & $15.98$  & $4.46$                          & $16.96$\\
$2.900\times 10^{-9}$   & $17.49$  &$9.98 \times 10^{-2}$  & $16.90$\\
$2.950\times 10^{-9}$   & $17.49$  &$9.98 \times 10^{-2}$  & $16.62$\\
$3.000\times 10^{-9}$   & $17.49$  &$9.96 \times 10^{-2}$  & $16.34$\\
$3.250\times 10^{-9}$   & $17.49$  &$9.93 \times 10^{-2}$  & $15.08$\\
$3.500\times 10^{-9}$   & $17.49$  &$9.92 \times 10^{-2}$  & $14.01$\\
$3.750\times 10^{-9}$   & $17.48$  &$9.99 \times 10^{-2}$  & $13.07$\\
$4.000\times 10^{-9}$   & $17.48$  &$1.01 \times 10^{-1}$  & $12.26$\\
$6.000\times 10^{-9}$   & $17.46$  &$9.94 \times 10^{-2}$  & $8.17$\\
$8.000\times 10^{-9}$   & $17.45$  &$9.69 \times 10^{-2}$  & $6.13$\\
$1.000\times 10^{-8}$   & $17.44$  &$9.32 \times 10^{-2}$  & $4.90$\\
$1.000\times 10^{-7}$   & $17.28$  &$9.21 \times 10^{-2}$  & $4.90\times 10^{-1}$\\
$1.000\times 10^{-6}$   & $17.28$  &$9.21 \times 10^{-2}$  & $4.90\times 10^{-2}$\\
$1.000\times 10^{-5}$   & $17.28$  &$9.21 \times 10^{-2}$  &  $4.90\times 10^{-3}$\\
\hline\hline
 \end{tabular}
\label{dapp}
\end{table}

It appears then that the spreading of the lifetime of the large drops could probably be justified in terms of a time-dependent protein adsorption. It must be kept in mind however, that in our simulations the same value of the surface excess is used for the drops and the interface, while in the experiment, it is only the approaching drops which are likely to be partially covered by the proteins.  

Ghosh and Juvekar \cite{ghosh:2002} published photographs which demonstrate that the collision of a millimetric-size ($r_i = 1.6\,\mathrm{mm}$) drop with a planar interface resembles the behavior of an under-damped system: ``the drop and the interface undergo an oscillatory (up-and-down) motion before attaining the rest position''. As discussed by these authors, elevations of the drop during this motion with respect to its final equilibrium position at rest, indicate that the movement is similar to that of a rubber ball dropped on a stretched membrane. Such behavior is very different from the over-damped motion predicted by film-thinning models.

In regard to the above observations, we studied the motion of the drops assuming a time-dependent surfactant adsorption but deactivating the mechanism of capillary waves. It was observed that:

\begin{description}
 
\item[a.] If a very small diffusion constant is used, $D_\mathrm{app} = 2.9 \times 10^{-12}$ m$^2$/s, the film thins smoothly until coalescence occurs.

\item [b.] If a very large diffusion constant is used, $D_\mathrm{app} = 2.9 \times 10^{-7}$ m$^2$/s, the film thins monotonically until it reaches $h = 29.6\,$nm. After this time ($t=1.19\,$s), the drop maintains its average distance from the interface fluctuating within $0.30\,$ nm.

\item[c.] For intermediate values of $D_\mathrm{app}$ like $2.9 \times 10^{-9}$ m$^2$/s, a drop of $r_i =200\,\mu$m reaches much smaller separations $h = 11.8\,$nm ($t=1.6\,$ s) before the repulsive barrier builds up. Next, the drop is pushed outward by the potential until it reaches $h = 29.7\,$nm. Following it oscillates around this distance. 

\item [d.] When the mechanism of capillary waves is activated for the case of intermediate apparent diffusion constants (i.e. $D_\mathrm{app} = 2.9 \times 10^{-9}$ m$^2$/s), the coalescence occurs during the outward motion of the drop. It does not happen at its closest separation from the surface.

\end{description}

From the above observations it appears likely that the bouncing of a drop at the oil/water interface is caused by the {\it sudden appearance} of a repulsive force at a short distance of approach. That repulsive force might occur as a result of a time-dependent adsorption, but it might also happen through the accumulation of surfactant at the so-called ``barrier ring'' as proposed by Ghosh and Juvekar \cite{ghosh:2002}.

\section{Conclusion}
\label{conclusions}

The lifetime of drops pressed by buoyancy against a planar interface can be reproduced using Emulsion Stability Simulations. Our calculations support previous theoretical evidence that suggests that the parti\-cular shape of the curve of $\tau$ vs. $r_i$ is basically determined by the drainage of the intervening film between the drops and the interface. However, depending on the particle size, different expressions of the initial distance of deformation and the maximum film radius of the drops are necessary. In the case of large drops the appropriate equations results from the consideration of the buoyancy force experienced by the drops. In the case of small non-deformable drops the (attractive) potential of interaction plays a significant role. The intermediate range of particle sizes between these two extremes results from a balance between interaction forces and the buoyancy force. Still, the calculations support the occurrence of deformable drops between $10\,\mu$m and $100\,\mu$m of particle radii. However, further study of this size range is necessary.

While the average value of the coalescence time can be justified by the mechanism of film thinning, the dispersion of the lifetime of large drops cannot be. As shown by the simulations, a possible explanation of this phenomenon comprises a combination of a substantial repulsive barrier for coalescence, the occurrence of capillary waves and a time-dependent surfactant adsorption.

\section*{Acknowledgements}

The suggestions and experimental data provided by Dr. T. D. Gurkov are gratefully acknowledged. Clara Rojas acknowledges the assistance of Dr. Aly J. Castellanos in the estimation of the Hamaker constant of soybean oil using the formalism of Lifshitz.

\bibliographystyle{apsrev}

\begin{thebibliography}{51}
\expandafter\ifx\csname natexlab\endcsname\relax\def\natexlab#1{#1}\fi
\expandafter\ifx\csname bibnamefont\endcsname\relax
  \def\bibnamefont#1{#1}\fi
\expandafter\ifx\csname bibfnamefont\endcsname\relax
  \def\bibfnamefont#1{#1}\fi
\expandafter\ifx\csname citenamefont\endcsname\relax
  \def\citenamefont#1{#1}\fi
\expandafter\ifx\csname url\endcsname\relax
  \def\url#1{\texttt{#1}}\fi
\expandafter\ifx\csname urlprefix\endcsname\relax\def\urlprefix{URL }\fi
\providecommand{\bibinfo}[2]{#2}
\providecommand{\eprint}[2][]{\url{#2}}

\bibitem[{\citenamefont{{E. Dickinson, B. S. Murray and G.
  Stainsby}}(1988)}]{dickinson:1988}
\bibinfo{author}{\bibnamefont{{E. Dickinson, B. S. Murray and G. Stainsby}}},
  \bibinfo{journal}{J. Chem. Soc., Faraday Trans. 1}
  \textbf{\bibinfo{volume}{84}}, \bibinfo{pages}{871} (\bibinfo{year}{1988}).

\bibitem[{\citenamefont{{E. S. Basheva, T. D. Gurkov, I. B. Ivanov, G. B.
  Bantchev and B. Campbell and R. P. Borwankar}}(1999)}]{basheva:1999}
\bibinfo{author}{\bibnamefont{{E. S. Basheva, T. D. Gurkov, I. B. Ivanov, G. B.
  Bantchev and B. Campbell and R. P. Borwankar}}}, \bibinfo{journal}{Langmuir}
  \textbf{\bibinfo{volume}{15}}, \bibinfo{pages}{6754} (\bibinfo{year}{1999}).

\bibitem[{\citenamefont{{L. D. Landau and E. M. Lifshitz}}(1984)}]{landau:1984}
\bibinfo{author}{\bibnamefont{{L. D. Landau and E. M. Lifshitz}}},
  \emph{\bibinfo{title}{Fluid Mechanics}} (\bibinfo{publisher}{Pergamon Press:
  Oxford}, \bibinfo{year}{1984}).

\bibitem[{\citenamefont{{A. Vrij and J. Th. G. Overbeek}}(1968)}]{vrij:1968}
\bibinfo{author}{\bibnamefont{{A. Vrij and J. Th. G. Overbeek}}},
  \bibinfo{journal}{J. Am. Chem. Soc.} \textbf{\bibinfo{volume}{90}},
  \bibinfo{pages}{3074} (\bibinfo{year}{1968}).

\bibitem[{\citenamefont{{A. Vrij}}(1966)}]{vrij:1966}
\bibinfo{author}{\bibnamefont{{A. Vrij}}}, \bibinfo{journal}{Discuss. Faraday
  Soc.} \textbf{\bibinfo{volume}{42}}, \bibinfo{pages}{23}
  (\bibinfo{year}{1966}).

\bibitem[{\citenamefont{{G. Urbina-Villalba and M.
  Garc\'ia-Sucre}}(2000)}]{urbina:2000}
\bibinfo{author}{\bibnamefont{{G. Urbina-Villalba and M. Garc\'ia-Sucre}}},
  \bibinfo{journal}{Langmuir} \textbf{\bibinfo{volume}{16}},
  \bibinfo{pages}{7975} (\bibinfo{year}{2000}).

\bibitem[{\citenamefont{{G. Urbina-Villalba, J. Toro-Mendoza, A. Lozs\'an and
  M. Garc\'ia-Sucre}}(2004)}]{urbina:2004}
\bibinfo{author}{\bibnamefont{{G. Urbina-Villalba, J. Toro-Mendoza, A. Lozs\'an
  and M. Garc\'ia-Sucre}}}, \emph{\bibinfo{title}{Emulsions: Structure
  Stability and Interactions}} (\bibinfo{publisher}{Elsevier},
  \bibinfo{year}{2004}), chap. \bibinfo{chapter}{Brownian Dynamics Simulation
  of Emulsion Stability}, pp. \bibinfo{pages}{677--719}.

\bibitem[{\citenamefont{{G. Urbina-Villalba}}(2009)}]{urbina:2009a}
\bibinfo{author}{\bibnamefont{{G. Urbina-Villalba}}}, \bibinfo{journal}{Int. J.
  Mol. Sci} \textbf{\bibinfo{volume}{10}}, \bibinfo{pages}{1}
  (\bibinfo{year}{2009}).

\bibitem[{\citenamefont{{J. Toro-Mendoza, A. Lozs\'an, M. Garc\'ia-Sucre, 
  A. J. Castellanos and G. Urbina-Villalba}}(2010)}]{toro:2010}
\bibinfo{author}{\bibnamefont{{J. Toro-Mendoza, A. Lozs\'an, M. Garc\'ia-Sucre,
   A. J. Castellanos  and G. Urbina-Villalba}}}, \bibinfo{journal}{Phys. Rev. E.}
  \textbf{\bibinfo{volume}{81}}, \bibinfo{pages}{011405}
  (\bibinfo{year}{2010}).

\bibitem[{\citenamefont{{C. Rojas, G. Urbina-Villalba and M.
  Garc\'ia-Sucre}}(2010)}]{rojas:2010a}
\bibinfo{author}{\bibnamefont{{C. Rojas, G. Urbina-Villalba and M.
  Garc\'ia-Sucre}}}, \bibinfo{journal}{Phys. Rev. E}
  \textbf{\bibinfo{volume}{81}}, \bibinfo{pages}{016302}
  (\bibinfo{year}{2010}).

\bibitem[{\citenamefont{{D. L. Ermak and J. A. McCammon}}(1978)}]{ermak:1978}
\bibinfo{author}{\bibnamefont{{D. L. Ermak and J. A. McCammon}}},
  \bibinfo{journal}{J. Chem. Phys.} \textbf{\bibinfo{volume}{69}},
  \bibinfo{pages}{1352} (\bibinfo{year}{1978}).

\bibitem[{\citenamefont{{I. B. Ivanov, K. D. Danov and P. A.
  Kralchevsky}}(1999)}]{ivanov:1999}
\bibinfo{author}{\bibnamefont{{I. B. Ivanov, K. D. Danov and P. A.
  Kralchevsky}}}, \bibinfo{journal}{Coll. Surf. A}
  \textbf{\bibinfo{volume}{152}}, \bibinfo{pages}{161} (\bibinfo{year}{1999}).

\bibitem[{\citenamefont{{K. D. Danov, N. D. Denkov, D. N. Petsev, I. B. Ivanov
  and R. Borwankar}}(1993)}]{danov:1993b}
\bibinfo{author}{\bibnamefont{{K. D. Danov, N. D. Denkov, D. N. Petsev, I. B.
  Ivanov and R. Borwankar}}}, \bibinfo{journal}{Langmuir}
  \textbf{\bibinfo{volume}{9}}, \bibinfo{pages}{1731} (\bibinfo{year}{1993}).

\bibitem[{\citenamefont{{K. D. Danov, D. N. Petsev, N. D. Denkov and R.
  Borwankar}}(1993)}]{danov:1993a}
\bibinfo{author}{\bibnamefont{{K. D. Danov, D. N. Petsev, N. D. Denkov and R.
  Borwankar}}}, \bibinfo{journal}{J. Chem. Phys.}
  \textbf{\bibinfo{volume}{99}}, \bibinfo{pages}{7179} (\bibinfo{year}{1993}).

\bibitem[{\citenamefont{{I. B. Ivanov, D. S. Dimitrov, P. Somasundaran and R.
  K. Jain}}(1985)}]{ivanov:1985}
\bibinfo{author}{\bibnamefont{{I. B. Ivanov, D. S. Dimitrov, P. Somasundaran
  and R. K. Jain}}}, \bibinfo{journal}{Chem. Eng. Sci.}
  \textbf{\bibinfo{volume}{40}}, \bibinfo{pages}{137} (\bibinfo{year}{1985}).

\bibitem[{\citenamefont{{A. Scheludko}}(1967)}]{scheludko:1967}
\bibinfo{author}{\bibnamefont{{A. Scheludko}}}, \bibinfo{journal}{Adv. Coll.
  Int. Sci.} \textbf{\bibinfo{volume}{1}}, \bibinfo{pages}{391}
  (\bibinfo{year}{1967}).

\bibitem[{\citenamefont{{I. B. Ivanov, B. Radoev, E. Manev and A.
  Scheludko}}(1970)}]{ivanov:1970}
\bibinfo{author}{\bibnamefont{{I. B. Ivanov, B. Radoev, E. Manev and A.
  Scheludko}}}, \bibinfo{journal}{Trans. Faraday Soc.}
  \textbf{\bibinfo{volume}{66}}, \bibinfo{pages}{1262} (\bibinfo{year}{1970}).

\bibitem[{\citenamefont{{E. D. Manev and A. V. Nguyen}}(2005)}]{manev:2005a}
\bibinfo{author}{\bibnamefont{{E. D. Manev and A. V. Nguyen}}},
  \bibinfo{journal}{Adv. Coll. Int. Sci.} \textbf{\bibinfo{volume}{114-115}},
  \bibinfo{pages}{133} (\bibinfo{year}{2005}).

\bibitem[{\citenamefont{{E. D. Manev and J. K. Angarska}}(2005)}]{manev:2005b}
\bibinfo{author}{\bibnamefont{{E. D. Manev and J. K. Angarska}}},
  \bibinfo{journal}{Coll. Surf. A} \textbf{\bibinfo{volume}{263}},
  \bibinfo{pages}{250} (\bibinfo{year}{2005}).

\bibitem[{\citenamefont{{A. Lozs\'an, M. Garc\'ia-Sucre and G.
  Urbina-Villalba}}(2006)}]{lozsan:2006}
\bibinfo{author}{\bibnamefont{{A. Lozs\'an, M. Garc\'ia-Sucre and G.
  Urbina-Villalba}}}, \bibinfo{journal}{J. Coll. Int. Sci}
  \textbf{\bibinfo{volume}{299}}, \bibinfo{pages}{366} (\bibinfo{year}{2006}).

\bibitem[{\citenamefont{{P. A. Kralchevsky, T. D. Gurkov and I. B.
  Ivanov}}(1991)}]{kralchevsky:1991a}
\bibinfo{author}{\bibnamefont{{P. A. Kralchevsky, T. D. Gurkov and I. B.
  Ivanov}}}, \bibinfo{journal}{Coll. Surf.} \textbf{\bibinfo{volume}{56}},
  \bibinfo{pages}{149} (\bibinfo{year}{1991}).

\bibitem[{\citenamefont{{P. A. Kralchevsky and T. D.
  Gurkov}}(1991)}]{kralchevsky:1991b}
\bibinfo{author}{\bibnamefont{{P. A. Kralchevsky and T. D. Gurkov}}},
  \bibinfo{journal}{Coll. Surf.} \textbf{\bibinfo{volume}{56}},
  \bibinfo{pages}{101} (\bibinfo{year}{1991}).

\bibitem[{\citenamefont{{H. C. Hamaker}}(1937)}]{hamaker:1937}
\bibinfo{author}{\bibnamefont{{H. C. Hamaker}}}, \bibinfo{journal}{Physica}
  \textbf{\bibinfo{volume}{IV}}, \bibinfo{pages}{1058} (\bibinfo{year}{1937}).

\bibitem[{\citenamefont{{A. Lozs\'an, M. Garc\'ia-Sucre and G.
  Urbina-Villalba}}(2005)}]{lozsan:2005}
\bibinfo{author}{\bibnamefont{{A. Lozs\'an, M. Garc\'ia-Sucre and G.
  Urbina-Villalba}}}, \bibinfo{journal}{Phys. Rev. E.}
  \textbf{\bibinfo{volume}{72}}, \bibinfo{pages}{061405}
  (\bibinfo{year}{2005}).

\bibitem[{\citenamefont{{T. D. Gurkov and E.S. Basheva}}(2002)}]{gurkov:2002}
\bibinfo{author}{\bibnamefont{{T. D. Gurkov and E.S. Basheva}}},
  \emph{\bibinfo{title}{Encyclopedia of surface and colloid science, Volume 4}}
  (\bibinfo{publisher}{Taylor \& Francis}, \bibinfo{year}{2002}), chap.
  \bibinfo{chapter}{Hydrodynamic Behavior and Stability of Approaching
  Deformable Drops}, p. \bibinfo{pages}{2773}.

\bibitem[{\citenamefont{{J.N. Israelachvili}}(1998)}]{israelachvili:1998}
\bibinfo{author}{\bibnamefont{{J.N. Israelachvili}}},
  \emph{\bibinfo{title}{Intermolecular and surface forces}}
  (\bibinfo{publisher}{Academic Press}, \bibinfo{year}{1998}),
  chap.~\bibinfo{chapter}{11}, p.~\bibinfo{pages}{90}.

\bibitem[{\citenamefont{{C. J. Drummond and D. Y. C.
  Chan}}(1996)}]{drummond:1996}
\bibinfo{author}{\bibnamefont{{C. J. Drummond and D. Y. C. Chan}}},
  \bibinfo{journal}{Langmuir} \textbf{\bibinfo{volume}{12}},
  \bibinfo{pages}{3356} (\bibinfo{year}{1996}).

\bibitem[{oli()}]{olive}
\urlprefix\url{http://en.wikipedia.org/wiki/Olive_oil}.

\bibitem[{\citenamefont{{L. Kong, J. K. Beattie and R. J.
  Hunter}}(2003)}]{kong:2003}
\bibinfo{author}{\bibnamefont{{L. Kong, J. K. Beattie and R. J. Hunter}}},
  \bibinfo{journal}{Coll. Surf. B} \textbf{\bibinfo{volume}{27}},
  \bibinfo{pages}{11} (\bibinfo{year}{2003}).

\bibitem[{\citenamefont{{N. J. Freeman, L. L. Peel, M. J Swann, G. H. Cross, A.
  Reeves, S. Brand and J. R. Lu}}(2004)}]{freeman:2004}
\bibinfo{author}{\bibnamefont{{N. J. Freeman, L. L. Peel, M. J Swann, G. H.
  Cross, A. Reeves, S. Brand and J. R. Lu}}}, \bibinfo{journal}{J. Phys.:
  Condens. Matter} \textbf{\bibinfo{volume}{16}}, \bibinfo{pages}{S2493}
  (\bibinfo{year}{2004}).

\bibitem[{\citenamefont{{Y. L. Jeyachandran, E. Mielczarski, B. Rai and J. A.
  Mielczarski}}(2009)}]{jeyachandran:2009}
\bibinfo{author}{\bibnamefont{{Y. L. Jeyachandran, E. Mielczarski, B. Rai and
  J. A. Mielczarski}}}, \bibinfo{journal}{Langmuir}
  \textbf{\bibinfo{volume}{25}}, \bibinfo{pages}{11614} (\bibinfo{year}{2009}).

\bibitem[{\citenamefont{{D. E. Graham and M. C. Phillips}}(1979)}]{graham:1979}
\bibinfo{author}{\bibnamefont{{D. E. Graham and M. C. Phillips}}},
  \bibinfo{journal}{J. Coll. Int. Sci.} \textbf{\bibinfo{volume}{70}},
  \bibinfo{pages}{415} (\bibinfo{year}{1979}).

\bibitem[{\citenamefont{{M. Tirado-Miranda, A. Schmitt, J. Callejas-Fern\'andez
  and A. Fern\'andez-Barbero}}(2003)}]{tirado:2003}
\bibinfo{author}{\bibnamefont{{M. Tirado-Miranda, A. Schmitt, J.
  Callejas-Fern\'andez and A. Fern\'andez-Barbero}}}, \bibinfo{journal}{J.
  Chem. Phys.} \textbf{\bibinfo{volume}{119}}, \bibinfo{pages}{9251}
  (\bibinfo{year}{2003}).

\bibitem[{\citenamefont{{C. Czeslik, G. Jackler, T. Hazlett, E. Gratton, R.
  Steitz, A. Wittemann and M. Ballauff}}(2004)}]{czeslik:2004}
\bibinfo{author}{\bibnamefont{{C. Czeslik, G. Jackler, T. Hazlett, E. Gratton,
  R. Steitz, A. Wittemann and M. Ballauff}}}, \bibinfo{journal}{Phys. Chem.
  Chem. Phys.} \textbf{\bibinfo{volume}{6}}, \bibinfo{pages}{5557}
  (\bibinfo{year}{2004}).

\bibitem[{\citenamefont{{I. B. Ivanov and P. A.
  Kralchevsky}}(1997)}]{ivanov:1997}
\bibinfo{author}{\bibnamefont{{I. B. Ivanov and P. A. Kralchevsky}}},
  \bibinfo{journal}{Coll. Surf. A} \textbf{\bibinfo{volume}{128}},
  \bibinfo{pages}{155} (\bibinfo{year}{1997}).

\bibitem[{\citenamefont{{A. Vrij}}(1964)}]{vrij:1964}
\bibinfo{author}{\bibnamefont{{A. Vrij}}}, \bibinfo{journal}{J. Coll. Int. Sci}
  \textbf{\bibinfo{volume}{19}}, \bibinfo{pages}{1} (\bibinfo{year}{1964}).

\bibitem[{\citenamefont{{G. D. M. Mackay and S. G. Mason}}(1963)}]{mackay:1963}
\bibinfo{author}{\bibnamefont{{G. D. M. Mackay and S. G. Mason}}},
  \bibinfo{journal}{Can. J. Chem. Eng.} \textbf{\bibinfo{volume}{41}},
  \bibinfo{pages}{203} (\bibinfo{year}{1963}).

\bibitem[{\citenamefont{{O. Reynolds}}(1881)}]{reynolds:1881}
\bibinfo{author}{\bibnamefont{{O. Reynolds}}}, \bibinfo{journal}{Chem. News}
  \textbf{\bibinfo{volume}{44}}, \bibinfo{pages}{211} (\bibinfo{year}{1881}).

\bibitem[{\citenamefont{{K. P. Velikov, O. D. Velev, K. G. Marinova and G. N.
  Constantinides}}(1997)}]{velikov:1997}
\bibinfo{author}{\bibnamefont{{K. P. Velikov, O. D. Velev, K. G. Marinova and
  G. N. Constantinides}}}, \bibinfo{journal}{J. Chem. Soc., Faraday Trans.}
  \textbf{\bibinfo{volume}{93}}, \bibinfo{pages}{2069} (\bibinfo{year}{1997}).

\bibitem[{\citenamefont{{A. Nikolova and D. Exerowa}}(1999)}]{nikolova:1999}
\bibinfo{author}{\bibnamefont{{A. Nikolova and D. Exerowa}}},
  \bibinfo{journal}{Coll. Surf. A} \textbf{\bibinfo{volume}{149}},
  \bibinfo{pages}{185} (\bibinfo{year}{1999}).

\bibitem[{\citenamefont{{D. Kashchiev and D. Exerowa}}(1980)}]{kashchiev:1980}
\bibinfo{author}{\bibnamefont{{D. Kashchiev and D. Exerowa}}},
  \bibinfo{journal}{J. Coll. Int. Sci} \textbf{\bibinfo{volume}{77}},
  \bibinfo{pages}{501} (\bibinfo{year}{1980}).

\bibitem[{\citenamefont{{P. Ghosh and V. A. Juvekar}}(2002)}]{ghosh:2002}
\bibinfo{author}{\bibnamefont{{P. Ghosh and V. A. Juvekar}}},
  \bibinfo{journal}{Trans. IChemE} \textbf{\bibinfo{volume}{80}},
  \bibinfo{pages}{715} (\bibinfo{year}{2002}).

\bibitem[{\citenamefont{{S. J. Alexander}}(1977)}]{alexander:1977}
\bibinfo{author}{\bibnamefont{{S. J. Alexander}}}, \bibinfo{journal}{Physique}
  \textbf{\bibinfo{volume}{38}}, \bibinfo{pages}{983} (\bibinfo{year}{1977}).

\bibitem[{\citenamefont{{P. G. de Gennes}}(1987)}]{degennes:1987}
\bibinfo{author}{\bibnamefont{{P. G. de Gennes}}}, \bibinfo{journal}{Adv. Coll.
  Int. Sci.} \textbf{\bibinfo{volume}{27}}, \bibinfo{pages}{189}
  (\bibinfo{year}{1987}).

\bibitem[{\citenamefont{Creighton}(1984)}]{creighton:1984}
\bibinfo{author}{\bibfnamefont{T.~E.} \bibnamefont{Creighton}},
  \emph{\bibinfo{title}{Proteins}} (\bibinfo{publisher}{W. H. Freeman and
  Company}, \bibinfo{year}{1984}).

\bibitem[{\citenamefont{{G. Urbina-Villalba}}(2004)}]{urbina:2004a}
\bibinfo{author}{\bibnamefont{{G. Urbina-Villalba}}},
  \bibinfo{journal}{Langmuir} \textbf{\bibinfo{volume}{20}},
  \bibinfo{pages}{3872} (\bibinfo{year}{2004}).

\bibitem[{\citenamefont{{L. Liggieri, F. Ravera and A.
  Passerone}}(1996)}]{liggieri:1996}
\bibinfo{author}{\bibnamefont{{L. Liggieri, F. Ravera and A. Passerone}}},
  \bibinfo{journal}{csA} \textbf{\bibinfo{volume}{114}}, \bibinfo{pages}{351}
  (\bibinfo{year}{1996}).

\bibitem[{\citenamefont{{A. F. H. Ward and L. J. Tordai}}(1946)}]{ward:1946}
\bibinfo{author}{\bibnamefont{{A. F. H. Ward and L. J. Tordai}}},
  \bibinfo{journal}{J. Chem. Phys.} \textbf{\bibinfo{volume}{14}},
  \bibinfo{pages}{453} (\bibinfo{year}{1946}).

\bibitem[{\citenamefont{{M. J. Rosen and X. Y. Hua}}(1990)}]{rosen:1990}
\bibinfo{author}{\bibnamefont{{M. J. Rosen and X. Y. Hua}}},
  \bibinfo{journal}{jcis} \textbf{\bibinfo{volume}{139}}, \bibinfo{pages}{397}
  (\bibinfo{year}{1990}).

\bibitem[{\citenamefont{{X. Y. Hua and M. J. Rosen}}(1998)}]{hua:1988}
\bibinfo{author}{\bibnamefont{{X. Y. Hua and M. J. Rosen}}},
  \bibinfo{journal}{jcis} \textbf{\bibinfo{volume}{124}}, \bibinfo{pages}{652}
  (\bibinfo{year}{1998}).

\bibitem[{\citenamefont{{X. Y. Hua and M. J. Rosen}}(1991)}]{hua:1991}
\bibinfo{author}{\bibnamefont{{X. Y. Hua and M. J. Rosen}}},
  \bibinfo{journal}{jcis} \textbf{\bibinfo{volume}{141}}, \bibinfo{pages}{180}
  (\bibinfo{year}{1991}).

\end{thebibliography}

\end{document}